\documentclass[fleqn,usenatbib]{mnras}

\usepackage{newtxtext,newtxmath}
\usepackage{comment}

\usepackage[T1]{fontenc}
\usepackage{ae,aecompl}
\usepackage{amsmath}
\usepackage{nccmath}

\usepackage{graphicx}	
\usepackage{amsmath}	
\usepackage{amssymb}	

\usepackage{soul} 

\title[Gone after one orbit]{Gone after one orbit: How cluster environments quench galaxies}

\author[Marcel Lotz et al.]{Marcel Lotz$^{1,2,3}$\thanks{E-mail: mlotz@usm.lmu.de},
Rhea-Silvia Remus$^{1,3}$,
Klaus Dolag$^{1,3,4}$,
Andrea Biviano$^{5,6}$, 
\newauthor and Andreas Burkert$^{1,2,3,7}$
\\
$^{1}$Universit\"{a}ts-Sternwarte M\"{u}nchen, Fakult\"{a}t f\"{u}r Physik, LMU Munich, Scheinerstr. 1, 81679 M\"{u}nchen, Germany\\
$^{2}$Excellence Cluster Universe, Boltzmannstra\ss{}e 2, 85748 Garching, Germany\\
$^{3}$Excellence Cluster ORIGINS, Boltzmannstra\ss{}e 2, 85748 Garching, Germany\\
$^{4}$Max-Planck Institute for Astrophysics, Karl-Schwarzschild-Stra\ss{}e 1, 85741 Garching, Germany\\
$^{5}$INAF-Osservatorio Astronomico di Trieste, via G.B. Tiepolo 11, 34131, Trieste, Italy\\
$^{6}$IFPU - Institute for Fundamental Physics of the Universe, via Beirut 2, 34014, Trieste, Italy\\
$^{7}$Max-Planck Institute for Extraterrestrial Physics, Giessenbachstra\ss{}e 1, 85748 Garching, Germany
}

\date{Accepted XXX. Received YYY; in original form ZZZ}

\pubyear{2019}

\begin{document}
\label{firstpage}
\pagerange{\pageref{firstpage}--\pageref{lastpage}}
\maketitle

\begin{abstract}

The effect of galactic orbits on a galaxy's internal evolution within a galaxy cluster environment has been the focus of heated debate in recent years.
To understand this connection, we use both the $(0.5 \,$Gpc)$^3$ and the Gpc$^3$ boxes from the cosmological hydrodynamical simulation set \textit{Magneticum Pathfinder}. We investigate the velocity-anisotropy, phase space, and the orbital evolution of up to $\sim 5 \cdot 10^{5}$ resolved satellite galaxies within our sample of 6776 clusters with $M_{\mathrm{vir}} > 10^{14} \, \mathrm{M_{\odot}}$ at low redshift, which we also trace back in time. In agreement with observations, we find that star-forming satellite galaxies inside galaxy clusters are characterised by more radially dominated orbits, independent of cluster mass. Furthermore, the vast majority of star-forming satellite galaxies stop forming stars during their first passage. We find a strong dichotomy both in line-of-sight and radial phase space between star-forming and quiescent galaxies, in line with observations. 
The tracking of individual orbits shows that the star-formation of almost all satellite galaxies drops to zero within $1 \, \mathrm{Gyr}$ after in-fall. Satellite galaxies that are able to remain star-forming longer are characterised by tangential orbits and high stellar mass. 
All this indicates that in galaxy clusters the dominant quenching mechanism is ram-pressure stripping.

\end{abstract}




\section{Introduction}

Already in 1901 \cite{1901AN....155..127W} noticed that the galaxy population in clusters is different from the population in the field. Later it was shown that a morphology-density relation for galaxies  exists \citep{1980ApJ...236..351D, 2003MNRAS.346..601G}. While late-type or irregular galaxies are most abundant in the low density `field', early-type galaxies are found in increasingly dense environments, i.e. groups and clusters. The strong implication is that a morphological transition from late- to early-type galaxies is driven by the increasing environmental density.

Cosmological simulations find that, independent of the original formation environment, galaxies continuously move towards higher density regions \citep{2005Natur.435..629S}. Considering these findings a picture of galaxy evolution emerges: galaxies tend towards higher density regions and thereby undergo internal and morphological evolution. This makes cluster environments an integral part of galaxy evolution, as the study provides insights into a fundamental and often terminal chapter of their existence \citep{1984ARA&A..22..185D}.

Galaxy clusters are the largest gravitationally bound structures in the observable Universe, consisting of hundreds of galaxies and, as such, offer a unique opportunity to study galaxy evolution \citep{2012ARA&A..50..353K}. They greatly impact the star-formation properties of galaxies in their vicinity \citep{2014A&ARv..22...74B}. Due to environmental quenching galaxies in clusters are, hence, far more likely to have reduced star-formation in comparison to field galaxies \citep{1974ApJ...194....1O, 1978ApJ...219...18B, 1980ApJ...236..351D, 1997ApJ...488L..75B}.
One of the aims of current galaxy cluster research is to understand, quantify, and determine the regions in which quenching mechanisms act. Observations have established that galaxy cluster environments extend out to $2-3 \, \mathrm{R_{vir}}$, much further than previously assumed \citep{2000ApJ...540..113B, 2002AJ....124.2440S, 2008A&A...486....9V, 2009A&A...500..947B, 2009ApJ...699.1333H, 2010MNRAS.404.1231V, 2012MNRAS.424..232W}.

Computational research suggests that star-formation quenching begins $2-3 \, \mathrm{Gyr}$ prior to entry of the inner cluster region \citep{2018MNRAS.475.3654Z}. This implies that galaxies are already partially quenched through halo gas removal in the outskirts $(2-3 \, \mathrm{R_{vir}})$ of the cluster while crossing the shock heated ($10^7 - 10^8 \, \mathrm{K}$), X-ray emitting accretion shock region \citep{1988xrec.book.....S}.
Several quenching mechanisms are under debate: relevant for the outskirts of clusters are \textit{strangulation}, i.e. stopping the replenishment of a galaxy's cold gas due to no longer being able to accrete gas within the cluster environment and \textit{pre-processing}, i.e. the removal of hot halo gas in a group or merger environment prior to cluster in-fall \citep{2012MNRAS.424.1179B}.
Once the in-falling galaxies enter the inner region $(0.5 \, \mathrm{R_{vir}})$, the removal of star-forming disc gas through ram-pressure stripping becomes efficient \citep{2018MNRAS.475.3654Z}.

Essentially, this means that the hot gas halo is removed prior to the cold gas within the galaxy \citep{1980ApJ...237..692L, 2008MNRAS.383..593M, 2009MNRAS.399.2221B, 2013MNRAS.430.3017B}. This suggests that the primary quenching mechanism at radii $r \gtrsim 0.5-1.0 \, \mathrm{r_{vir}}$ is strangulation or pre-processing. Further, this implies that outside the inner cluster region the cold gas is not necessarily expelled or significantly heated, but rather the supply in the halo is affected and thus star-formation dwindles due to a lack of fresh halo gas \citep{2008ApJ...672L.103K}. This model of galaxy quenching in and around clusters would imply a gradual decrease in star-formation in the cluster outskirts, followed by a rapid shutdown in the inner regions.

Throughout this paper we use different scaling radii, velocities and masses. To avoid confusion, we clarify the different definitions used: when solely presenting results from our simulation we scale according to the virial radius $r_{\mathrm{vir}}$, virial velocity $v_{\mathrm{vir}}$, and virial mass $M_{\mathrm{vir}}$. The virial radius is calculated via a top-hat spherical collapse model \citep{1996MNRAS.282..263E}.
When comparing to observations, we reproduce the prescription the observations use. In practice there are two relevant scaling definitions: Firstly, we use $R_{\mathrm{200,crit}}$, i.e. the radius encompassing an over-density 200 times larger than the critical density, $v_{\mathrm{200,crit}}$, and $M_{\mathrm{200,crit}}$.
Secondly, we use $R_{\mathrm{200,mean}}$, $v_{\mathrm{200,mean}}$, and $M_{\mathrm{200,mean}}$, which are defined analogous to above but using the mean background matter density rather than the critical density. To avoid confusion between 2D and 3D radii, we use $R$ to indicate the former and $r$ to indicate the latter.

To summarise, the main goal of this paper is to study different mechanisms which quench satellite galaxies in clusters. In Section \ref{sec:data}, we present the \textit{Magneticum Pathfinder} simulations and the relevant comparison observations and surveys. In Section \ref{sec:aniso}, we analyse the velocity-anisotropy profiles of satellite galaxies in comparison to observations and dependent on mass and redshift. In Section \ref{sec:PS}, we study the line-of-sight (LOS) space diagrams and compare to observations. Thereafter, we study the radial phase space diagrams and their dependence on mass and redshift. In Section \ref{sec:track}, we track the orbits of satellite galaxies through the simulation and look at the responsible quenching mechanisms. 
Finally, in Section \ref{sec:disc}, we discuss our findings and the relevance of ram-pressure stripping. In Section \ref{sec:conc}, we present our conclusions.

\section{Data sample}
\label{sec:data}

\subsection{Magneticum Pathfinder simulations}
\textit{Magneticum Pathfinder} is a set of large scale smoothed-particle hydrodynamic (SPH) simulations that employ a mesh-free Lagrangian method aimed at following structure formation on cosmological scales. The boxes utilised during the scope of this paper are Box2/hr ($352 \, \mathrm{(Mpc/h)^3}$) and Box2b/hr ($640 \, \mathrm{(Mpc/h)^3}$). Box2/hr has a higher temporal resolution, i.e. a greater number of individual \texttt{SUBFIND} halo finder outputs \citep{2001NewA....6...79S, 2009MNRAS.399..497D}. Hence, it allows for the temporal tracking of subhalos rather than solely considering individual static redshifts.
In contrast, Box2b/hr is larger and thus allows for a more extended statistical sample, especially with regard to high mass clusters. 

Box2/hr is comprised of $2 \cdot 1584^3$ particles, while Box2b/hr is comprised of $2 \cdot 2880^3$ particles. Both boxes resolution level is set to `high resolution' (\textit{hr}): dark matter (dm) and gas particles have masses of $m_{\mathrm{dm}} = 6.9 \cdot 10^8 \, h^{-1}\mathrm{M_{\odot}}$ and $m_{\mathrm{gas}} = 1.4 \cdot 10^8 \, h^{-1}\mathrm{M_{\odot}}$, respectively. Stellar particles are formed from gas particles and have $\sim 1/4$ of the mass of their parent gas particle. At this resolution level the softening of the dark matter, gas and stars is $\epsilon_{\mathrm{dm}} = 3.75 \, h^{-1} \mathrm{kpc}$, $\epsilon_{\mathrm{gas}} = 3.75 \, h^{-1} \mathrm{kpc}$ and $\epsilon_{\mathrm{stars}} = 2 \, h^{-1} \mathrm{kpc}$, respectively.
These two boxes have already been used in the past to study various galaxy cluster properties: the ICL-BCG connection in comparison to observations \citep{2017Galax...5...49R}, the effect of baryons on large scale structure \citep{2016MNRAS.456.2361B, 2018MNRAS.478.1305C}, the intra-cluster-medium (ICM) and inter-galactic medium (IGM) \citep{2015MNRAS.451.4277D, 2017Galax...5...35D, 2018arXiv180401096B} and active galactic nuclei (AGN) \citep{2014MNRAS.442.2304H, 2018MNRAS.481..341S}.

The astrophysical processes modelled within the Magneticum simulation include, but are not limited to: cooling, star-formation and winds \citep{2003MNRAS.339..289S}, metals, stellar populations and chemical enrichment from AGB stars \citep{2003MNRAS.342.1025T, 2006JON.....5..858T, 2017Galax...5...35D}, black holes and AGN feedback \citep{2014MNRAS.442.2304H}, thermal conduction \citep{2004ApJ...606L..97D}, low viscosity scheme to track turbulence \citep{2005MNRAS.364..753D,2016MNRAS.455.2110B}, higher order SPH kernels \citep{2012MNRAS.425.1068D} and magnetic fields (passive) \citep{2009MNRAS.398.1678D}. For a more in-depth appreciation of the precise physical processes refer to \cite{2009MNRAS.399..497D} and \cite{2015ApJ...812...29T}.

For both boxes and all redshifts satellite galaxies are selected to have a minimum stellar mass of $M_* = 3.5 \cdot 10^9 \, h^{-1} \mathrm{M_{\odot}}$, corresponding to a minimum of $\sim 100$ stellar particles for a given galaxy. We choose this threshold since it implies that the galaxy has had enough gas to successfully produce a minimum stellar component.

In order to differentiate between star-forming and quiescent satellite galaxies, the criterion introduced by \cite{2008ApJ...688..770F} is used throughout this paper at all redshifts: the measure for star-formation is given by $\mathrm{SSFR} \cdot t_{\mathrm{H}}$, i.e. the specific star-formation rate $\mathrm{SSFR}$ multiplied by the Hubble time $t_{\mathrm{H}}$. Galaxies with a value above $\mathrm{SSFR} \cdot t_{\mathrm{H}} > 0.3$ are classified as star-forming, while galaxies with $\mathrm{SSFR} \cdot t_{\mathrm{H}} < 0.3$ are classified as quiescent. 

Importantly, this `blueness criterion' ($\mathrm{SSFR} \cdot t_{\mathrm{H}} > 0.3$) is time dependent rather than merely being applicable to low redshifts. Hence, this definition encompasses the changing star-formation history on a cosmological scale and is well suited for a temporal comparison. With this criterion, the Milky Way would have $\mathrm{SSFR} \cdot t_{\mathrm{H}} \sim 0.4$ at $z = 0$ and, hence, be considered star-forming \citep{2015ApJ...806...96L}.

Throughout this paper the following cosmology is adopted: $h = 0.704$, $\Omega_M = 0.272$, $\Omega_{\Lambda} = 0.728$ and $\Omega_b = 0.0451$. As the simulation outputs are independent of little h, we re-calibrated the simulations to the little h value of the observations in question when comparing our results.

\subsection{Observational comparison with CLASH}
\label{sec:obs}

The CLASH survey observes 25 massive galaxy clusters with the Hubble Space Telescope's panchromatic imaging equipment (Wide-field Camera 3, WFC3, and the Advanced Camera for Surveys, ACS) \citep{2012ApJS..199...25P, 2014egcc.confE..12M}. One of four primary science goals of CLASH is the study of internal structure and galaxy evolution within and behind the clusters. Thus, it provides an interesting data set to compare with the Magneticum simulation. The CLASH cluster under consideration (MACS J1206.2-0847) is located at a redshift of $z = 0.44$. The cluster hosts $590$ identified members obtained at the ESO VLT \citep{2014Msngr.158...48R} and has a mass of $M_{200} = (1.4 \pm 0.2) \cdot 10^{15} \, \mathrm{M_{\odot}}$  with a concentration of $c_{200} = 6 \pm 1$ \citep{2013A&A...558A...1B}. The data was obtained by the VLT/VIMOS large program, which aims at constraining the cluster mass profile over the radial range of $0$-$2.5$ virial radii.

Studies regarding the velocity-anisotropy include \cite{2009A&A...501..419B, 2014A&A...566A..68M, 2016A&A...585A.160A, 2016A&A...594A..51B}. However, they did not investigate the difference between star-forming and quiescent galaxies at the level of detail as demonstrated by \cite{2013A&A...558A...1B}, especially regarding the confidence regions of the populations. The cluster ${\mathrm{MACS J1206.2-0847}}$ is to date the only CLASH cluster for which the velocity-anisotropy was calculated for both passive and star-forming galaxies. The velocity-anisotropy profiles calculated in \cite{2016A&A...594A..51B} from the GCLASS survey at $z \sim 1$ have such large confidence regions that a wide range of profiles are encompassed by said regions.

\section{Velocity-anisotropy Profiles} 
\label{sec:aniso}

The goal of parametrising and comparing objects through the use of the velocity-anisotropy $\beta$ is to gain an understanding of the relative importance of different degrees of freedom, i.e. comparing the preference for tangentially versus radially dominated orbits. 
The premise being that if a population is dominated by tangential or radial movement the dispersion in the respective degree of freedom will be larger.
The velocity-anisotropy is defined as follows \citep{1987gady.book.....B}:

\begin{ceqn}
\begin{align}
 \beta \equiv 1 - \frac{\sigma_t^2}{\sigma_r^2} \mathrm{,}
\end{align}
\end{ceqn}

\begin{ceqn}
\begin{align}
\mathrm{with} \texttt{ } \sigma_t^2 = \frac{ \sigma_{\theta}^2+\sigma_{\phi}^2}{2} .
\end{align}
\end{ceqn}

\noindent
where $\sigma_r$, $\sigma_t$, $\sigma_{\theta}$ and $\sigma_{\phi}$ parametrise the velocity dispersions in the radial, tangential, $\theta$ and $\phi$ direction respectively. 
The degree of (an)isotropy is studied providing information on whether the galaxies are moving on radial ($\beta > 0$), tangential ($\beta < 0$) or isotropic ($\beta = 0$) orbits.

The velocity-anisotropy parameter $\beta$ is binned into equal distance radial differential bins. It is calculated for the total, the star-forming, and the quiescent satellite galaxy population separately. The error associated with the calculation of $\beta$ is derived via bootstrapping and error bars correspond to the $1 \sigma$ confidence interval.

\subsection{CLASH comparison}
\label{sub:anisoCLASH}

The velocity-anisotropy, when obtained via observations, is calculated through the inversion of the Jeans equation. This is motivated by the fact that velocities and, thus, dispersions can only be observed along the line-of-sight (LOS). For the $\beta (r)$ and LOS comparison between Magneticum and the CLASH cluster ${\mathrm{MACS J1206.2-0847}}$, simulated clusters are selected above a mass threshold of $M_{\mathrm{200,crit}} > 5 \cdot 10^{14} \mathrm{M_{\odot}}$. In order to provide a large statistical sample, both Box2/hr and Box2b/hr were taken into account. We use only clusters at a redshift of $z = 0.44$, resulting in 15 clusters for Box2/hr and 71 for Box2b/hr. The velocity-anisotropy comparison is done by considering a sphere around Magneticum clusters, whereas the LOS phase space comparison in Section \ref{sec:PS} considers a cylinder of height $89.4 \mathrm{Mpc}$, corresponding to the velocity range of cluster members \citep{2013A&A...558A...1B}.
To increase the statistical sample of the LOS phase space analysis, we considered the projection along three independent spatial axes. 

\begin{figure}
	\includegraphics[width=\columnwidth]{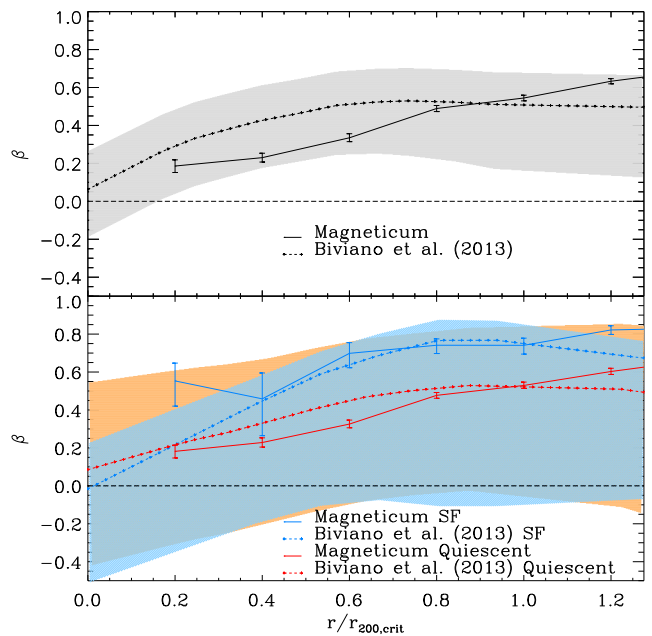}
    \caption{Box2/hr and Box2b/hr differential 3D radial profiles of velocity-anisotropy parameter at $ z = 0.44$ of clusters with mass threshold ${M_{\mathrm{200,crit}} > 5 \cdot 10^{14} \, \mathrm{M_{\odot}}}$. Solid lines describe the selected clusters from the Magneticum simulation, while the dashed lines describe the CLASH observations extracted from \protect\cite{2013A&A...558A...1B}. Top panel: total galaxy population profile. Bottom panel: star-forming (blue) and quiescent (red) population profiles. Shaded regions correspond to the $1 \sigma$ confidence zone of the respective observations. The horizontal dashed line indicates an isotropic velocity distribution. Values greater than $\beta = 0 $ correspond to more radial orbits, while negative values represent tangentially dominated orbits.}
    \label{fig:BetaBiv}
\end{figure}

Figure \ref{fig:BetaBiv} shows the radial $\beta$-profiles for the stacked Magneticum clusters and the CLASH cluster MACS J1206.2-0847. The top panel depicts the total population both from Magneticum (solid line) and \cite{2013A&A...558A...1B} (dashed line). The bottom panel displays the quiescent (red) and star-forming (blue) populations for Magneticum (solid lines) and the CLASH cluster (dashed lines), with the shaded region representing the $1 \sigma$ confidence zone of the observations. The vast majority of points lie within each other's $1 \sigma$ error regions and bars, respectively. In addition, the global behaviour of the observations is reproduced by the Magneticum profiles for both the star-forming and quiescent populations.
Considering that the observations are based on a single cluster with $590$ members, while Magneticum utilises $86$ clusters to derive its statistics at $z = 0.44$, the results are in good agreement. 

As can be clearly seen in Figure \ref{fig:BetaBiv} from both simulations and observations, the anisotropy profile increases with increasing radius. This indicates that the outskirts are more radially dominated than the inner regions of the clusters.
\cite{2012MNRAS.427.1024I} show that of all the galaxies entering a galaxy cluster at a certain time, those that will survive until redshift zero are those characterised by the most tangential orbits at in-fall. This indicates that longer galaxy cluster membership increases the likelihood of circular obits. 
Given that smaller radii are associated with a higher probability of long-term cluster membership, the anisotropy profile decreases towards smaller radii.

Intuitively one might assume that the accretion of radially in-falling galaxies leads to more radially dominated orbits of the total population with time, rather than the identified tendency towards isotropy. However, the in-falling population makes up less and less of the total population as time proceeds. 
In addition, the most radially dominated in-falling galaxies are likely to experience strong tidal stripping by the BCG, thus falling below our resolution limit and consequently no longer contributing to the $\beta (r)$ profiles \citep{2016A&A...585A.160A}.

Figure \ref{fig:BetaBiv} also clearly shows a difference between the star-forming and quiescent population. The star-forming population is more radially dominated than its quiescent counterpart. This suggests that the star-forming population is more in-fall dominated, as will be further shown in Section \ref{sub:PSmasstempevol}.
Subsequently, it is likely that the star-forming galaxies belong to a younger population of a given cluster, since they are more removed from isotropy ($\beta =0$).

Figure \ref{fig:BetaBiv} also shows that the total population (black solid line) is dominated by the quiescent population (red solid line). This is accentuated by the fact that the black and red lines are exceedingly similar. Only a small sample of galaxies are star-forming, with the bulk of clusters being made up of quiescent galaxies, with regard to our star-formation criterion.

The main source of error of the observational calculations are the uncertainties associated with the line-of-sight dispersion. The observational confidence zones depicted in Figure \ref{fig:BetaBiv} are calculated through modifications of the beta profile. These modifications are then inverted to yield a wide grid of predicted line-of-sight velocity dispersion profiles \citep{1994MNRAS.270..271V}. This reversed method is employed because error propagation through the Jeans inversion is infeasible \citep{2013A&A...558A...1B}. 

Although the observational method to evaluate $\beta (r)$ is sophisticated, issues remain. For one, the entire calculation of the velocity-anisotropy profiles hinges on the fiducial mass profile. The NFW profile, which was assumed and fitted, can be the source of large errors. Depending on the true nature of the mass profile, the NFW fit by \cite{2012ApJ...755...56U} does not necessarily fit well along the entire profile. This is especially significant at large and small radii, where the deviations from the NFW profile are likely to be largest. The error on quantities like the concentration $c_{200} = 6 \pm 1$ of the observed cluster are large and, hence, the fitted NFW profile could easily not accurately describe the actual mass profile \citep{2013A&A...558A...1B}. 

Additionally, interlopers have a potentially major impact on the observationally based calculations of the anisotropy profile. Given solely the line-of-sight velocity, it is virtually impossible to conclusively attribute a galaxy to an individual radial bin of a cluster. Subsequently, the calculation of the observed anisotropy profile will always be subject to inaccuracies. For example, the galaxies in the CLASH cluster (MACS J1206.2-0847) compared to in Figure \ref{fig:BetaBiv} are not distributed spatially isotropically \citep{2012ApJ...755...56U}. This impedes the identification of interlopers further, making their effect less predictable. In contrast, the Magneticum profiles depicted in Figure \ref{fig:BetaBiv} are the result of the construction of an exact sphere, which evaluates each radial shell. As expected by construction, when using the Clean membership identification algorithm \citep{2013MNRAS.429.3079M}, which has been used for several observational data-sets already (e.g. \cite{2013A&A...558A...1B, 2017A&A...607A..81B, 2017A&A...606A.108C}), we found a negligible impact by interlopers on our results.

In addition, ambiguity in the definition of the star-formation criterion gives rise to different categorisations of galaxies. The observers utilise colour-colour diagrams to identify star-forming and quiescent galaxies \citep{2014egcc.confE..12M}. In contrast, the criterion applied in the data reduction of the simulation considers the specific star-formation rate in dependence of redshift. This, however, does not account for recently formed stars, which impact the classification on the colour-colour diagram. As such, our star-formation criterion might underestimate the amount of star-forming galaxies compared to observations. As a result, quenching timescales are likely underestimated, too.

While observations of the CLASH cluster and the results from the Magneticum simulations are in very good agreement, there exist both simulations and observations that report opposite trends for quiescent and star-forming satellite populations: Both the Millennium Simulation \citep{2012MNRAS.427.1024I} and observations of the galaxy cluster Abell 85 \citep{2017MNRAS.468..364A} find that the quiescent population is characterised by more radially dominated orbits relative to the star-forming population. However, the Millennium Simulation is a DM-only simulation.
When searching for clusters with radially dominated orbits for quiescent satellites within the Magneticum simulation, we find a small number of clusters. Specifically, only $28$ of the $495$ most massive clusters ($h^{-1} M_{200,crit} > 10^{14} \, \mathrm{M_{\odot}}$), i.e. $5.7$ per cent in Box2/hr, were identified as having a more radially dominated quiescent population than was the case for the star-forming population. However, the vast majority of these clusters have extremely low (star-forming) galaxy numbers. We conclude that clusters matching the behaviour reported by \cite{2012MNRAS.427.1024I} and observed by \cite{2017MNRAS.468..364A} are rare in the Magneticum simulation, albeit they do exist.

\subsection{Quenching fractions}

\begin{figure}
	\includegraphics[width=\columnwidth]{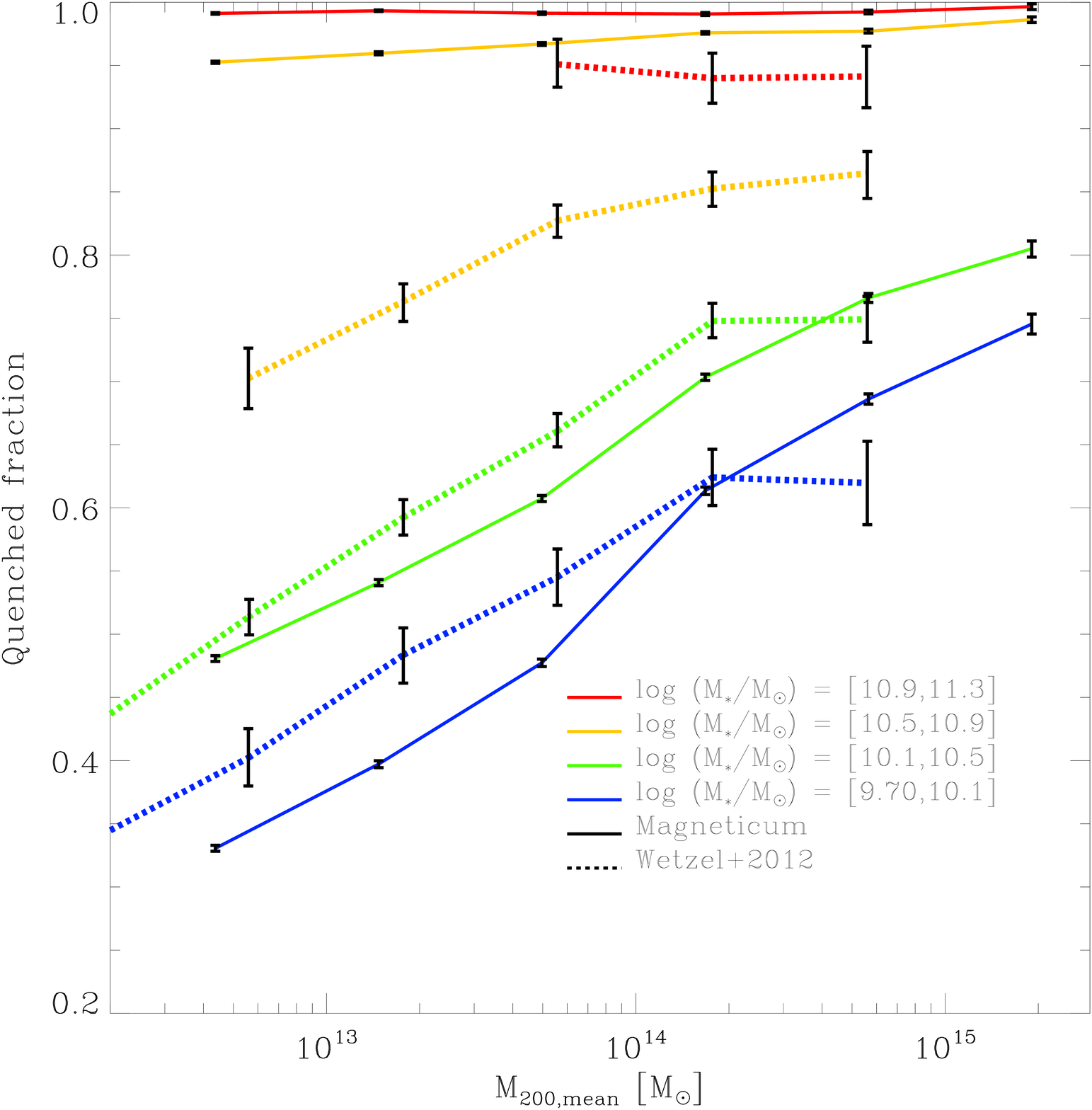}
    \caption{Box2/hr quenched satellite galaxy fractions in dependence of host halo mass, split into four stellar mass bins at $ z \sim 0$. Solid lines show the different stellar mass bins from the Magneticum simulation, while the dashed lines show the observations extracted from \protect\cite{2012MNRAS.424..232W}. The Magneticum errors are calculated via bootstrapping.}
    \label{fig:qfrac}
\end{figure}

To investigate the difference between star-forming and quiescent satellite galaxies further, we explore the impact of stellar mass and host halo mass on the fraction of quenched satellite galaxies. The quenched fraction is defined as the number of quenched satellite galaxies over the total number of satellite galaxies.
Figure \ref{fig:qfrac} shows the quenched fraction of satellite galaxies split into four different stellar mass intervals in the range of $\mathrm{log} (M_*/\mathrm{M_{\odot}}) = [9.7,11.3]$ in dependence of their host halo mass at $z = 0.066$. It compares the Box2/hr results (solid lines) to observations by \cite{2012MNRAS.424..232W} (dashed lines) at $z = 0.045$. 

\cite{2012MNRAS.424..232W} use galaxy group/cluster catalogues from the Sloan Digital Sky Survey Data Release 7 (SDSS DR7) with an overall median redshift of $z = 0.045$. \cite{2012MNRAS.424..232W} consider ${\sim 19000}$, $2200$, $160$ groups with ${M_{\mathrm{200,mean}} > 10^{12}, 10^{13}, 10^{14} \, \mathrm{M_{\odot}}}$, respectively, up to the most massive group at ${10^{15} \, \mathrm{M_{\odot}}}$ hosting $269$ satellites. The employed catalogue allows the examination of satellites with good statistics in the stellar mass range ${M_{*} = 5 \cdot 10^{9} - 2 \cdot 10^{11} \, \mathrm{M_{\odot}}}$.

By considering the stellar mass evolution at fixed halo mass and and the halo mass evolution at fixed stellar mass Figure \ref{fig:qfrac} offers a number of insights into environmental quenching dependencies: Firstly, the Magneticum simulation matches the observed increase in the fraction of quenched satellite galaxies with host halo mass, accurately describing the environmental impact on quenching. Secondly, it also reproduces an increase in quenched fraction with increasing stellar mass, accurately describing the impact of galactic feedback mechanisms on star-formation. Thirdly, a difference in the fraction of quenched satellite galaxies in the stellar mass range of $\mathrm{log} (M_{*}/\mathrm{M_{\odot}}) = [10.5,10.9]$ exists between Magneticum and the observations by \cite{2012MNRAS.424..232W}. Although there exists the discrepancy in the intermediate to high stellar mass range, overall the behaviour is in agreement.

The gap between the two stellar mass ranges ${\mathrm{log} (M_{*}/\mathrm{M_{\odot}}) = [10.1,10.5]}$ (green) and ${\mathrm{log} (M_{*}/\mathrm{M_{\odot}}) = [10.5,10.9]}$ (orange) suggests that the quenching of satellite galaxies in the Magneticum simulations is characterised by a strong stellar mass bimodality. Further it suggests that if the stellar mass is below ${\mathrm{log} (M_{*}/\mathrm{M_{\odot}}) \sim 10.5}$ satellite galaxies have a far lower probability of being quenched, while the opposite is true above this stellar mass threshold. The bimodality is strongest at lower host halo masses, implying that the quenching is not a result of the environment, but rather a self-regulatory behaviour, i.e. AGN feedback kicking in at a specific stellar mass as dictated by the implemented feedback model.

In contrast to the Magneticum results, the observations find a flattening in the stellar mass range of ${\mathrm{log} (M_{*}/\mathrm{M_{\odot}}) = [9.7,10.5]}$ at high host halo masses. The flattening of the observations is likely driven by the incompleteness of the SDSS in the high halo mass regime. As the SDSS median redshift is $z=0.045$, high host halo mass objects are rare \citep{2012MNRAS.424..232W}. Consequently, the Magneticum results provide a far more complete sample, especially in the high host halo mass regime compared to the observations.

Figure \ref{fig:qfrac} also displays the increasing impact of environmental quenching on lower stellar mass satellite galaxies. Below $\mathrm{log} (M_{*}/\mathrm{M_{\odot}}) \sim 10.5$ the quenched fraction of Magneticum results increases strongly with host halo mass, implying a strong environmental dependence. In the higher stellar mass range, $\mathrm{log} (M_{*}/\mathrm{M_{\odot}}) \gtrsim 10.5$, only a small host halo mass dependence is found. This is likely due to high stellar mass halos being quenched by feedback processes triggered by the strong mass growth, while low stellar mass satellites are more vulnerable to environmental quenching, lacking strong galactic feedback mechanisms inhibiting star-formation. A similar stellar mass dependence is found in the observations by \cite{2012MNRAS.424..232W}, however, the critical stellar mass threshold is $\mathrm{log} (M_{*}/\mathrm{M_{\odot}}) \sim 10.9$.

As described in Section \ref{sub:anisoCLASH}, we expect Magneticum to overestimate the amount of quenched satellite galaxies due to the blueness criterion evaluating the SSFR rather than the colour of satellite galaxies. In contrast, \cite{2012MNRAS.424..232W} find that an average of $10$ per cent of galaxies have been misassigned and state that the primary resultant bias is an underestimation of the quenched fraction for satellites. Therefore, a part of the discrepancy in Figure \ref{fig:qfrac} is likely the result of an overestimation on part of the simulation and an underestimation on part of the observations.

\subsection{Mass and temporal evolution}
\label{sub:betamasstemp}

\begin{figure*}
	\includegraphics[width=2.0\columnwidth]{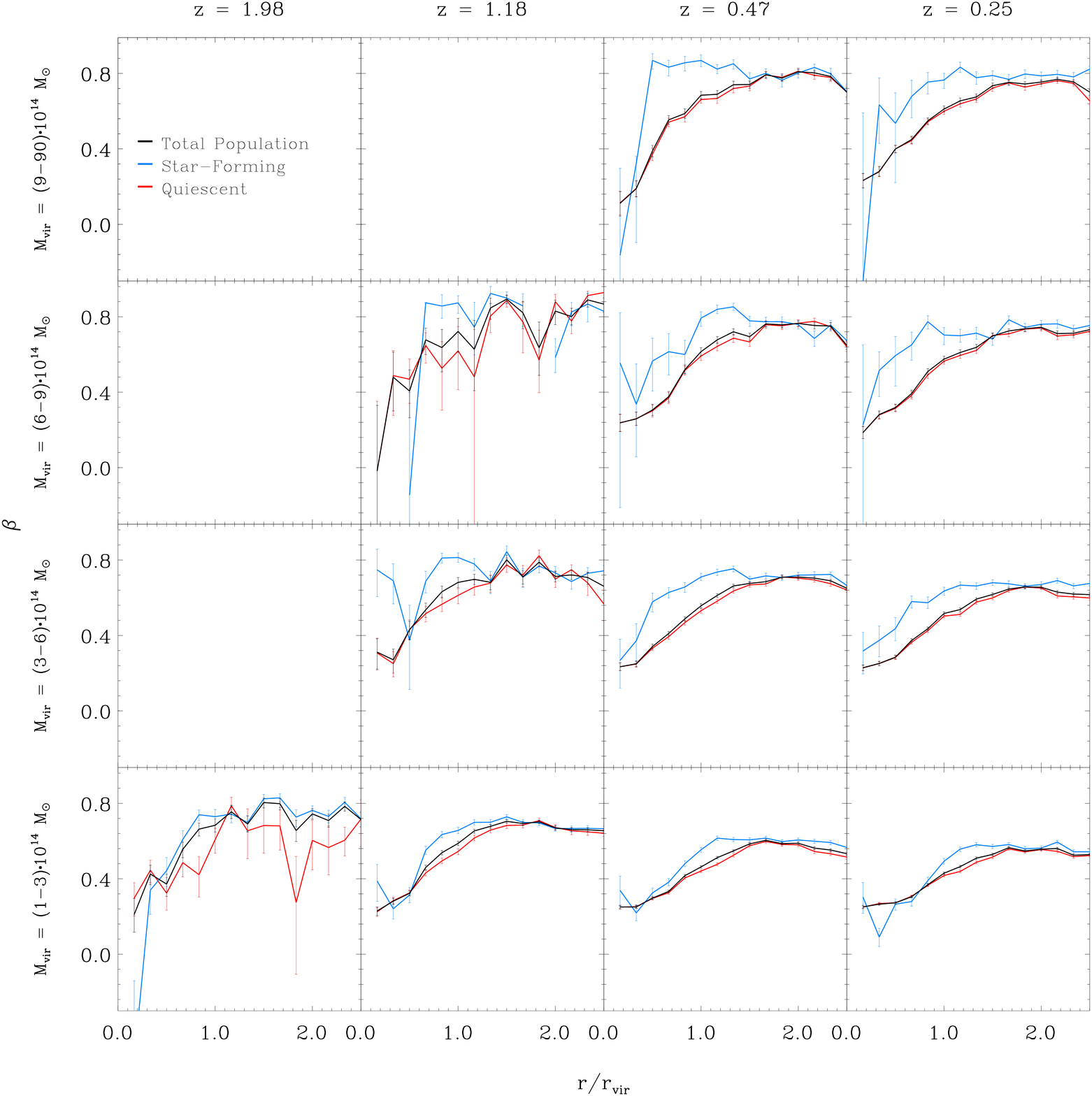}
    \caption{Combined Box2/hr and Box2b/hr differential 3D radial profiles of velocity-anisotropy parameter in dependence of redshift and virial cluster mass. Columns from left to right have the following redshifts: $z=1.98$, $z=1.18$, $z=0.47$ and $z=0.25$. Rows from top to bottom have the following cluster masses: $M_{\mathrm{vir}} = 9 - 90 \cdot 10^{14} \, \mathrm{M_{\odot}}$, $M_{\mathrm{vir}} = 6 - 9 \cdot 10^{14} \, \mathrm{M_{\odot}}$, $M_{\mathrm{vir}} = 3 - 6 \cdot 10^{14} \, \mathrm{M_{\odot}}$ and $M_{\mathrm{vir}} = 1 - 3 \cdot 10^{14} \, \mathrm{M_{\odot}}$. The blue lines correspond to the star-forming population, while the red lines correspond to the quiescent population. The total galaxy population is indicated by the black line. The empty panels are the result of a lack of clusters in the given redshift and cluster mass range.} 
    \label{fig:BetaGrid}
\end{figure*}

While we, so far, have focused on a comparison to observations, we now use the full predictive power of our simulations to study the velocity-anisotropy for clusters of different mass ranges through redshift. In Figure \ref{fig:BetaGrid} we evaluate the combined results from both Box2/hr and Box2b/hr. At the lowest redshift considered, $z = 0.25$, we calculate the velocity-anisotropy profiles of $\sim 5 \cdot 10^{5}$ resolved satellite galaxies within our sample of 6776 clusters with ${M_{\mathrm{vir}} > 10^{14} \, \mathrm{M_{\odot}}}$.

Figure \ref{fig:BetaGrid} shows two major similarities independent of mass and redshift. Firstly, the general behaviour of the profiles, albeit displaying differences, is fairly similar in all panels. Specifically, this entails that the star-forming population consistently lies on more radial orbits than its quiescent counterpart. Secondly, all profiles agree in their broad shape and trend towards isotropy ($\beta = 0$) at smaller radii. 

The outskirts are consistently radially dominated since at large radii the likelihood of galaxies being accreted, rather than already belonging to the cluster, is significantly higher. In addition, this appears to remain to hold true for the star-forming population at smaller radii too, i.e. the star-forming population is typically more in-fall dominated than the quiescent population.

Despite close similarities between panels in Figure \ref{fig:BetaGrid}, mass and redshift dependent trends can also be seen. Profiles clearly tend towards being more radially dominated with increasing cluster mass, as mass accretion is expected to occur progressively along small filaments, extending radially outside massive clusters \citep{2012MNRAS.427.1024I}. 
This results in a stronger signal towards radial accretion in the high mass end. In this picture, high mass clusters display more direct in-fall, i.e. radially dominated orbits, than low mass clusters.

Figure \ref{fig:BetaGrid} shows that the difference between the star-forming (blue) and quiescent (red) profiles becomes more pronounced with cluster mass. 
In the highest cluster mass bin ($M_{\mathrm{vir}} = 9 - 90 \cdot 10^{14} \, \mathrm{M_{\odot}}$) the star-forming anisotropy profile experiences a strong decline at $r/r_{\mathrm{vir}} < 0.5$. This is likely a result of the radially dominated star-forming galaxies being quenched more efficiently through ram-pressure stripping than the subset of more tangentially dominated star-forming galaxies. Hence, we see a selection mechanism, which preferentially quenches radially dominated orbits. 
This results in only the most circular star-forming galaxies existing at small radii, since the more radially dominated subset no longer exists, i.e. was quenched. Although this is most strongly exhibited in high mass clusters, most star-forming profiles exhibit a stronger decrease at smaller radii $r/r_{\mathrm{vir}} < 0.5 - 1$ than the quiescent population.
However, the remaining innermost $r/r_{\mathrm{vir}} < 0.5$ star-forming population is significantly smaller than the population at $r/r_{\mathrm{vir}} \sim 1$, calling for a higher number statistics investigation.

The only apparent redshift-dependent trend is that with decreasing redshift, less satellite galaxies are star-forming. This is visible from the fact that the red and black profiles are closer together at lower redshifts, i.e. the fraction of star-forming galaxies decreases. At high redshift ($z \sim 2$), the majority of galaxies are star-forming. This is not the case at any other redshift surveyed in Figure~\ref{fig:BetaGrid}. This implies that at high redshift more star-forming galaxies are accreted onto clusters, only to be quenched and become part of the low-redshift quiescent population.

We find that Figure \ref{fig:BetaGrid} is in good agreement with available observations. Specifically, the qualitative behaviour of the results from \cite{2016A&A...594A..51B} 
\footnote{The 10 stacked observational clusters in \cite{2016A&A...594A..51B} have a mass of $M_{200} = 4.5 \pm 1.2 \cdot 10^{14} \, \mathrm{M_{\odot}}$, which corresponds to a virial mass in the Magneticum simulation of $M_{\mathrm{vir}} \sim 3.8 \pm 1.0 \cdot 10^{14} \, \mathrm{M_{\odot}}$, considering that $M_{\mathrm{vir}}$ masses in Magneticum are typically a factor $\sim 1.2$ smaller at the considered redshift than the $M_{200}$ masses.} 
are well reproduced when considering the $z \sim 1$ column and the lower mass bins in Figure \ref{fig:BetaGrid}. 
Our results are also in agreement with \cite{2011MNRAS.412...49W}, who finds that satellite orbits are more radial and plunge deeper into their host haloes at higher host halo masses.

\section{Phase Space}
\label{sec:PS}

\begin{figure*}
	\includegraphics[width=\columnwidth]{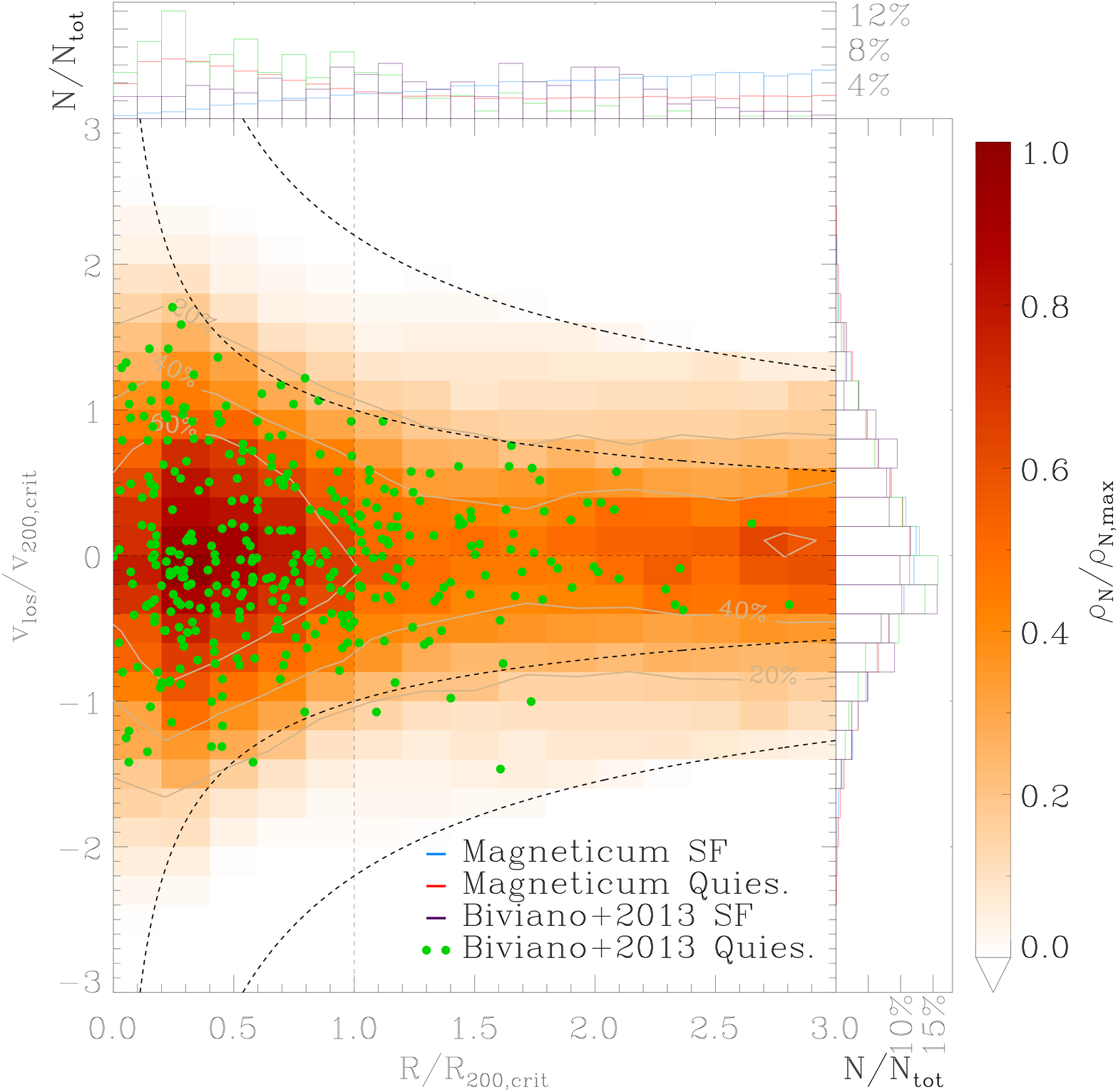}
    \includegraphics[width=\columnwidth]{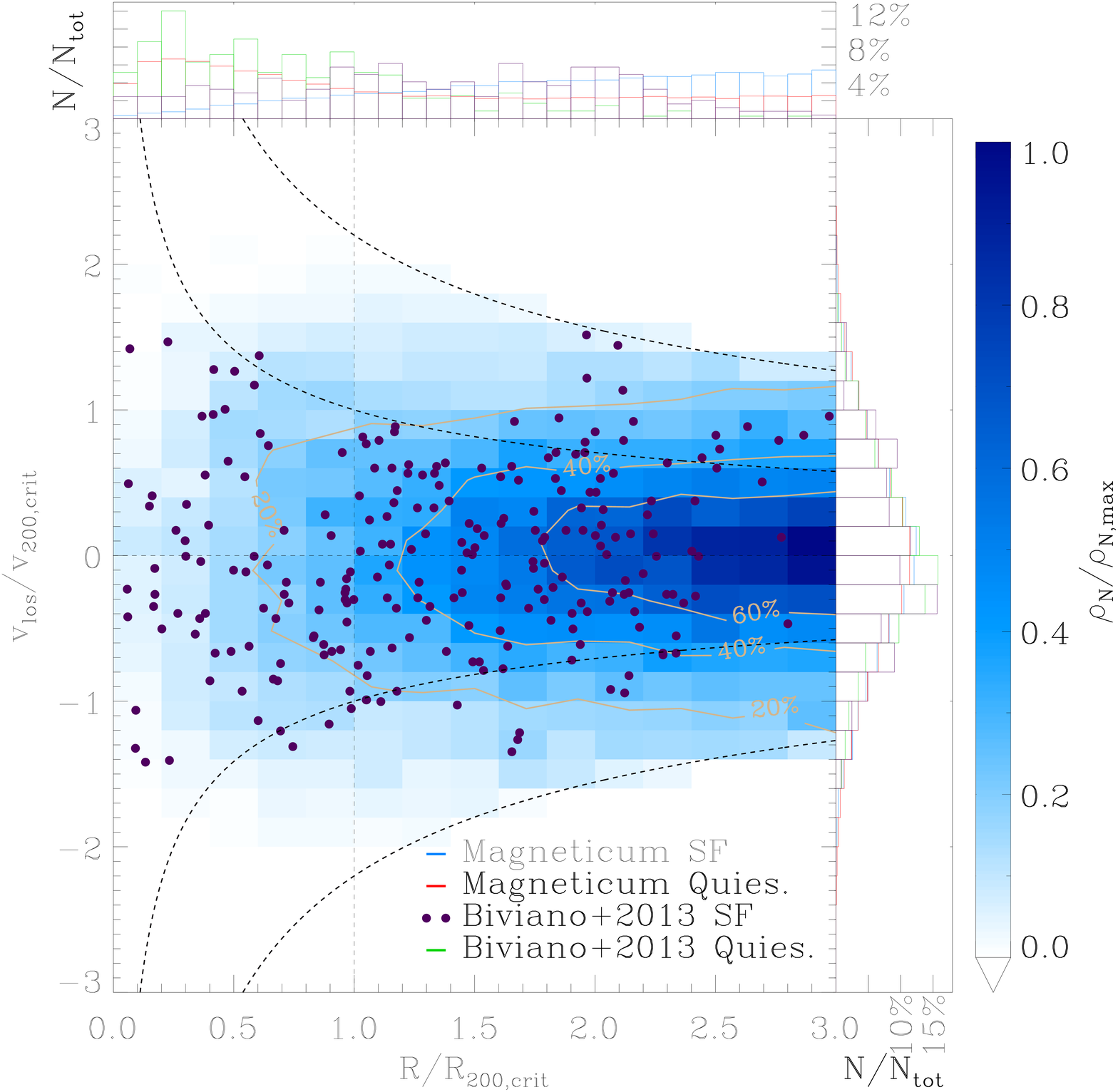}
    \caption{Combined Box2/hr and Box2b/hr quiescent (left) and star-forming (right) line-of-sight (LOS) phase space number density comparison with \protect\cite{2013A&A...558A...1B} and Magneticum clusters with a mass threshold of $M_{\mathrm{200,crit}} > 5 \cdot 10^{14} \, \mathrm{M_{\odot}}$ at $z = 0.44$. The figure displays the normalised LOS, i.e. the LOS velocity divided by the $v_{\mathrm{200,crit}}$ velocity ($v_{\mathrm{rad}}/v_{\mathrm{200,crit}}$), in dependence of the projected $R/R_{\mathrm{200,crit}}$ radius. The galaxies obtained through observations are characterised by green and purple indicating quiescent and star-forming galaxies, respectively. In addition, the histograms depict the relative abundance of each population projected onto the respective axes. Contour lines indicate the $20$, $40$, and $60$ per cent regions, with regard to the maximum quiescent LOS phase space density. The two sets of enveloping black lines correspond to $|v_{\mathrm{LOS}}/v_{\mathrm{200,crit}}| \sim |({R/R_{\mathrm{200,crit}}})^{-1/2}|$ and $|v_{\mathrm{LOS}}/v_{\mathrm{200,crit}}| \sim 2.2|({R/R_{\mathrm{200,crit}}})^{-1/2}|$.}
    \label{fig:ProjPS}
\end{figure*}

\subsection{Line-of-sight phase space}
\label{sub:LOS_PS}

Figure \ref{fig:ProjPS} shows the line-of-sight (LOS) phase space diagrams for the quiescent (left) and star-forming (right) population at $z=0.44$. The densities in each plot are scaled to the respective maximum density of the given population under consideration.
In both figures only satellite galaxies are included, i.e. brightest-cluster-galaxies (BCGs) are excluded. 
The enveloping dashed black lines in Figure~\ref{fig:ProjPS} are introduced to provide a relationship between the velocity and the radius via $|v_{\mathrm{LOS}}/v_{\mathrm{200,crit}}| \sim |({R/R_{\mathrm{200,crit}}})^{-1/2}|$. 
The basis of this relation is the $v \propto r^{-1/2}$ relation, which is derived from the virial theorem. The outer enveloping dashed black lines are motivated by the strongest outlier of the \citep{2013A&A...558A...1B} data. We exclude galaxies with properties that are not represented in the observations, as they are likely interlopers, i.e. galaxies that only lie within the cluster due to the LOS projection. We choose the following relation for the filtration: $|v_{\mathrm{LOS}}/v_{\mathrm{200,crit}}| \sim 2.2|({R/R_{\mathrm{200,crit}}})^{-1/2}|$, i.e. all galaxies larger than said relation are excluded.
The inner enveloping black line depicts the simple virial proportionality. Thus, it provides a theoretical description of the extent of the clusters.

The comparison between the two panels in Figure \ref{fig:ProjPS} demonstrates the complementary behaviour of the star-forming and quiescent galaxy populations and their individual preference for distinct regions in phase space. This preference, reflected in the phase space number density, is in agreement with the observed CLASH cluster galaxies \citep{2013A&A...558A...1B}. Typically, quiescent galaxies are located at radii $R < 1 \, \mathrm{R_{200,crit}}$, while star-forming galaxies have a significantly increased likelihood of being found in the outskirts of the cluster ($R > 1.5 \, \mathrm{R_{200,crit}}$).

At first glance, the left panel of Figure \ref{fig:ProjPS} seems to display a slight dichotomy in the Magneticum quiescent population. On the one hand we regard the overwhelming majority of the quiescent population at radii $R < 1 \, \mathrm{R_{200,crit}}$. On the other hand a second smaller population exists at $R \gtrsim 2.3 \, \mathrm{R_{200,crit}}$. However, if we consider the radial histogram (red), we are presented with a constant quiescent number density throughout both the intermediate region ($R \sim (1.2-2.3) \, \mathrm{R_{200,crit}}$) and the second smaller population ($R \gtrsim 2.3 \, \mathrm{R_{200,crit}}$). The apparent dichotomy and the constant radial histogram are likely the result of a mixture of two effects. Firstly, accretion in the outskirts, which is discussed more extensively in Section \ref{sub:PSmasstempevol}, drives the consistency in the radial histogram.
Secondly, the phase space density increases in the outskirts since the radial histogram remains constant, while the phase space volume decreases, as $|v_{\mathrm{LOS}}/v_{\mathrm{200,crit}}| \sim |({R/R_{\mathrm{200,crit}}})^{-1/2}|$ decreases.
We investigated whether the apparent over-density in the outskirts might be due to backsplash galaxies, however, we found no evidence to support this (also see Figure \ref{fig:PSGrid_r}).

The radial histograms, especially with regard to the star-forming population, also show good agreement between the simulation and the CLASH observations (blue and green lines in Figure~\ref{fig:ProjPS}): Both show a decrease below $R < 1 \, \mathrm{R_{200,crit}}$, hinting at a good description of the underlying quenching mechanism by the simulation. In the outskirts of the cluster ($R > 2.2 \, \mathrm{R_{200,crit}}$), the simulation no longer matches the observations. 
This is likely due to the CLASH cluster not accreting as many galaxies, both star-forming and quiescent, in the outskirts as is the case on average over a wide statistical sample. 

Possible other reasons for the different abundances of star-forming and quiescent galaxies at different radii, especially in the outskirts, include shocks. Shocks violently trigger star-formation only to form almost no stars after the initial star-burst. Galaxies, which have recently experienced star-bursts, thus look blue in colour observations, although technically they have a star-formation rate close to zero. As such, they are classified to be star-forming in \cite{2013A&A...558A...1B}, but are considered quiescent in the simulation which classifies according to the specific star-formation rate rather than the colour. This leads to an underestimation in Magneticum and an overestimation of the number of star-forming galaxies in observations.

To better understand whether the distribution of the galaxies extracted from the simulations within the line-of-sight phase space diagram (as shown in Figure \ref{fig:ProjPS}) is compatible with the distribution of observed galaxies, we computed the normalised distance between the cumulative distributions, as usually done when performing a Kolmogorov-Smirnov (KS) test. As it is not very meaningful to compute the probability from the KS test when comparing a single observed cluster with a large number of stacked clusters from the simulations, we compare this result to what we obtained via cluster by cluster variation from the 86 clusters extracted from the Magneticum simulation, see Figure \ref{fig:KShisto}. Here we compared the radial distribution as well as the line-of-sight distributions for star-forming and quiescent galaxies separately. Interestingly, we find that in the majority of cases the normalised distance of the cumulative distributions between the single cluster observation by \cite{2013A&A...558A...1B} and the stacked Magneticum clusters is much less than the expected cluster to cluster variation in the simulations. Even the result for the radial distribution of the star-forming galaxies, which shows by far the largest distance, is well within the 2 $\sigma$ region of the cluster by cluster variation. Within the limits of the observational data, we therefore can quantitatively confirm the impression resulting from Figure \ref{fig:ProjPS} that the distribution of galaxies within the simulations well represents the observed distribution within the line-of-sight phase space diagram.

\subsection{Mass and temporal evolution}
\label{sub:PSmasstempevol}

\begin{figure*}
	\includegraphics[width=2.0\columnwidth]{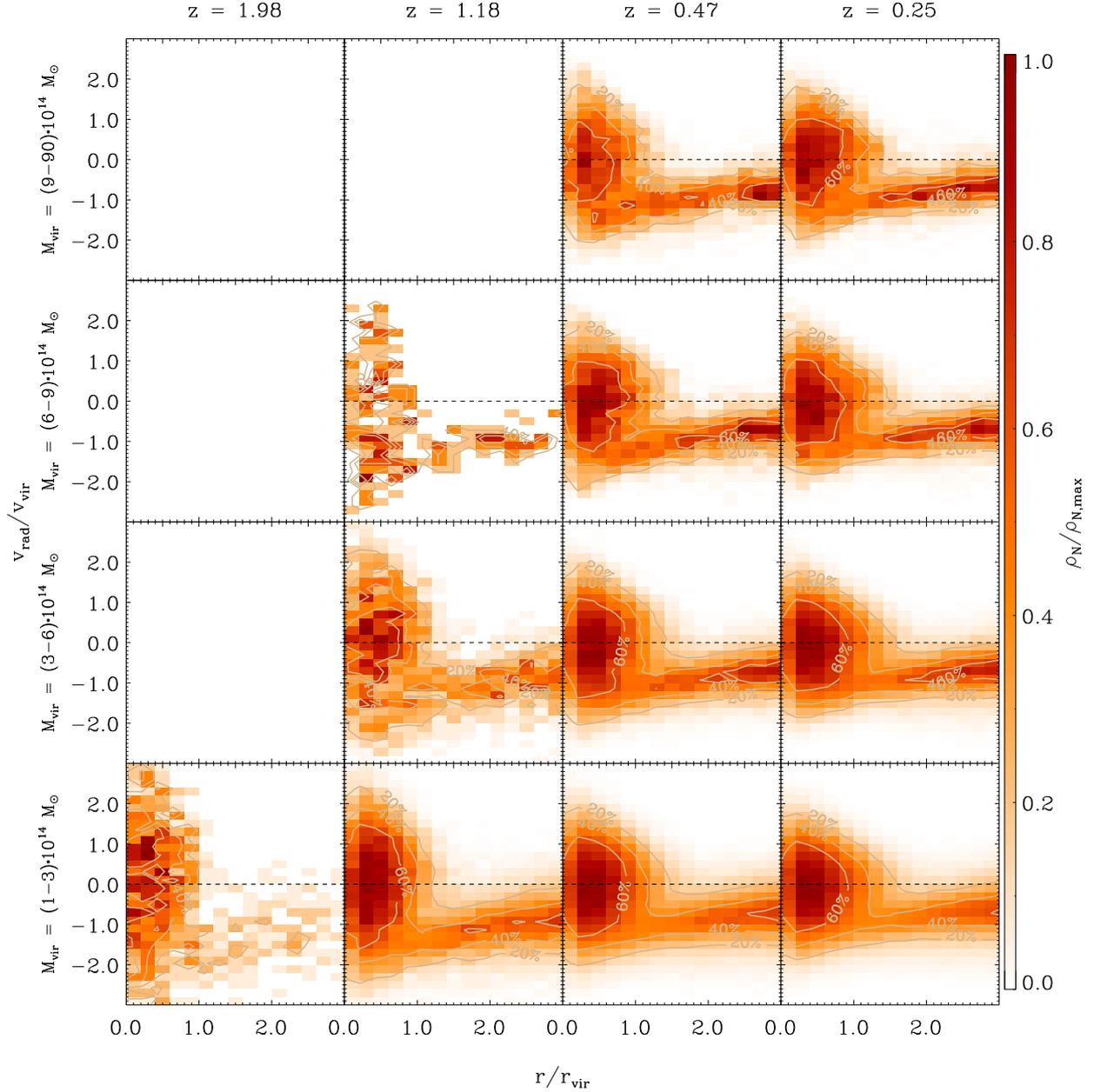}
    \caption{Combined Box2/hr and Box2b/hr normalised relative quiescent radial phase space number density as a function of the 3D radial profile in dependence of redshift and cluster mass. Columns from left to right have the following redshifts: $z=1.98$, $z=1.18$, $z=0.47$ and $z=0.25$. Rows from top to bottom have the following cluster masses: $M_{\mathrm{vir}} = 9 - 90 \cdot 10^{14} \, \mathrm{M_{\odot}}$, $M_{\mathrm{vir}} = 6 - 9 \cdot 10^{14} \, \mathrm{M_{\odot}}$, $M_{\mathrm{vir}} = 3 - 6 \cdot 10^{14} \, \mathrm{M_{\odot}}$ and $M_{\mathrm{vir}} = 1 - 3 \cdot 10^{14} \, \mathrm{M_{\odot}}$. The colourbar displays the relative phase space number density normalised to the maximum value of each individual panel. The contour lines correspond to 20, 40, and 60 per cent of the maximum density in each panel. The empty panels are the result of a lack of clusters in the given redshift and cluster mass range.}
    \label{fig:PSGrid_r}
\end{figure*}

\begin{figure*}
	\includegraphics[width=2.0\columnwidth]{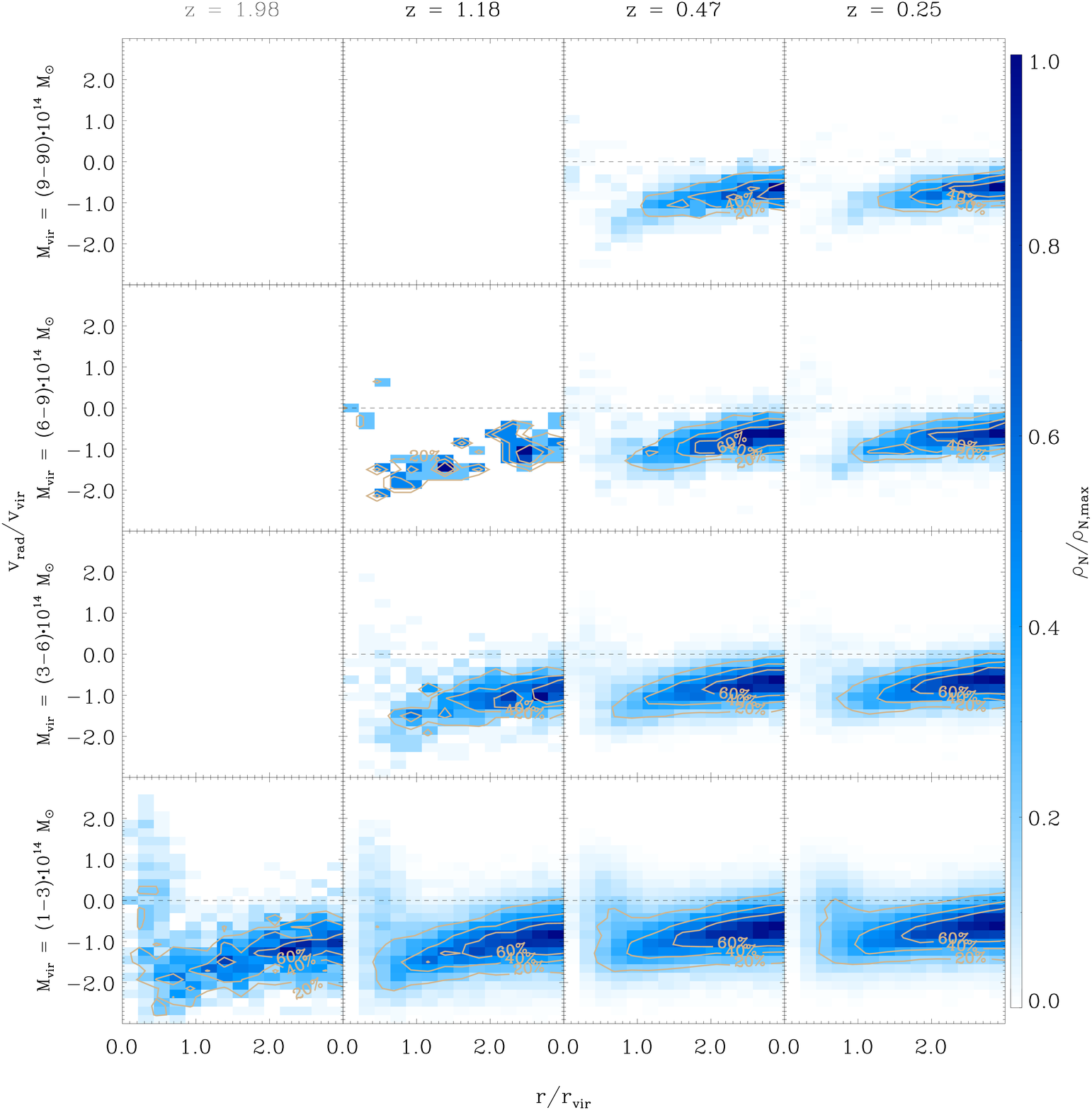}
    \caption{Combined Box2/hr and Box2b/hr normalised relative star-forming radial phase space number density as a function of the 3D radial profile in dependence of redshift and virial cluster mass. Columns from left to right have the following redshifts: $z=1.98$, $z=1.18$, $z=0.47$ and $z=0.25$. Rows from top to bottom have the following cluster masses: $M_{\mathrm{vir}} = 9 - 90 \cdot 10^{14} \, \mathrm{M_{\odot}}$, $M_{\mathrm{vir}} = 6 - 9 \cdot 10^{14} \, \mathrm{M_{\odot}}$, $M_{\mathrm{vir}} = 3 - 6 \cdot 10^{14} \, \mathrm{M_{\odot}}$ and $M_{\mathrm{vir}} = 1 - 3 \cdot 10^{14} \, \mathrm{M_{\odot}}$. The colourbar displays the relative phase space number density normalised to the maximum value of each individual panel. The contour lines correspond to 20, 40, and 60 per cent of the maximum density in each panel. The empty panels are the result of a lack of clusters in the given redshift and cluster mass range.}
    \label{fig:PSGrid_b}
\end{figure*}

Figures \ref{fig:PSGrid_r} and \ref{fig:PSGrid_b} show the temporal and cluster mass evolution of the quiescent and star-forming galaxy population in radial phase space density as a function of the radial profile. At the lowest redshift considered, $z = 0.25$, we calculate the phase space density of ${\sim 5 \cdot 10^{5}}$ resolved satellite galaxies within our sample of 6776 clusters with $M_{\mathrm{vir}} > 10^{14} \, \mathrm{M_{\odot}}$.
The contour lines correspond to the 20, 40, and 60 per cent thresholds of the maximum density in each panel. Both figures only display satellite galaxies and not BCGs.
Each panel in Figures \ref{fig:PSGrid_r} and \ref{fig:PSGrid_b} is normalised to the maximum value of its individual panel. Consequently, we compare relative densities rather than absolute phase space densities.
This relative normalisation is performed to better highlight the effect of quenching, rather than visualising changes in absolute phase space densities. As a result Figures \ref{fig:PSGrid_r} and \ref{fig:PSGrid_b} disclose no information about the absolute numbers of galaxies.

Considering the quiescent and star-forming population in Figures \ref{fig:PSGrid_r} and \ref{fig:PSGrid_b}, the following details are apparent. Firstly, the star-forming population is overwhelmingly dominated by in-fall. Specifically, the vast majority of star-forming galaxies are quenched outside $r > 0.5 \, \mathrm{r_{vir}}$ during their first passage, independent of redshift. Secondly, the long term cluster population is almost exclusively made up of quiescent galaxies. This is in agreement with previous hydrodynamic simulations and observations stating that the central cluster region is dominated by quiescent galaxies \citep{2006MNRAS.373..397S}. 
Thirdly, Figure~\ref{fig:PSGrid_r} suggests that recently accreted satellite galaxies are virialised, i.e. the system has reached virial equilibrium, during the first passage, as they cannot be distinguished from the virialised population. Specifically, there are almost no satellites with positive radial velocity at radii ${r \gtrsim 1.5 \, \mathrm{r_{vir}}}$. This is likely the result of a combination of virialisation and a scaling effect driven by the cluster growth which increases $\mathrm{r_{vir}}$, i.e. decreasing ${r/\mathrm{r_{vir}}}$ of a given orbit as the cluster accretes more mass.

Figure \ref{fig:PSGrid_r} demonstrates that high mass clusters exhibit more extensive accretion than low mass clusters, as implied by the higher relative density in the outskirts ($r > 2.0 \, \mathrm{r_{vir}}$).
The increased accretion in high mass clusters further strengthens the idea that high mass clusters are located within a surrounding high density environment, i.e. at the cross-section of filaments.

This is in agreement with previous simulations that show that the cluster environment, characterised by increased filamentary thickness, extends out to three to four times the virial radius \citep{2006MNRAS.370..656D}. 
Along filaments, the radial density profiles decrease less strongly than within the galaxy cluster \citep{2006MNRAS.370..656D}. 
This physical extension of the cluster environment manifests itself, amongst other things, through the decrease in the percentage of star-forming compared to quiescent galaxies with increasing cluster mass. The percentage of star-forming galaxies compared to their quiescent/passive counterparts (${N_{\mathrm{SF}}/N_{\mathrm{PAS}}}$) in cluster intervals ${M_{\mathrm{vir}} = 9 - 90 \cdot 10^{14} \, \mathrm{M_{\odot}}}$, ${M_{\mathrm{vir}} = 6 - 9 \cdot 10^{14} \, \mathrm{M_{\odot}}}$, ${M_{\mathrm{vir}} = 3 - 6 \cdot 10^{14} \, \mathrm{M_{\odot}}}$ and ${M_{\mathrm{vir}} = 1 - 3 \cdot 10^{14} \, \mathrm{M_{\odot}}}$ is ${N_{\mathrm{SF}}/N_{\mathrm{PAS}} = 8.9}$ per cent, ${N_{\mathrm{SF}}/N_{\mathrm{PAS}} = 10.4}$ per cent, ${N_{\mathrm{SF}}/N_{\mathrm{PAS}} = 12.0}$ per cent, and ${N_{\mathrm{SF}}/N_{\mathrm{PAS}} = 16.1}$ per cent, respectively. This implies that the overall environment of more massive clusters is more efficient in quenching, despite it extending out further than lower mass clusters.
In contrast, \cite{2013MNRAS.432..336W} find no dependence of host halo mass on quenching time-scales. This is likely due to the quenching in the Magneticum simulations being more strongly correlated with in-fall than is the case in the `delayed-then-rapid' quenching scenario found in \cite{2013MNRAS.432..336W}. Due to the long time delay, compared to the quenching time-scale, in the `delay-then-rapid' quenching scenario, host halo mass differences do not factor in as strongly when solely considering quenching time-scales.

Figure \ref{fig:PSGrid_b} offers more insights with regard to different mass bins and their temporal evolution: Lower cluster mass intervals present a lower quenching effectiveness than higher mass clusters. 
This is in agreement with observations, which find that processes responsible for the termination of star-formation in galaxy clusters are more effective in denser environments \citep{2012A&A...543A..19R}.
In the low cluster mass range more star-forming galaxies exist at smaller radii. In contrast to the higher cluster mass intervals, the low mass range also exhibits a star-forming, albeit small, population with positive radial velocity, especially at high redshift, indicating that quenching becomes more effective at lower redshift and higher cluster mass. This is in agreement with a suite of recent cluster resimulations by \cite{2019MNRAS.484.3968A}, which find ram-pressure stripping of subhalos to be significantly enhanced in more massive halos compared to less massive halos.

\subsection{Stellar mass comparison}
\label{sub:PS_stellar}

\begin{figure*}
	\includegraphics[width=2.0\columnwidth]{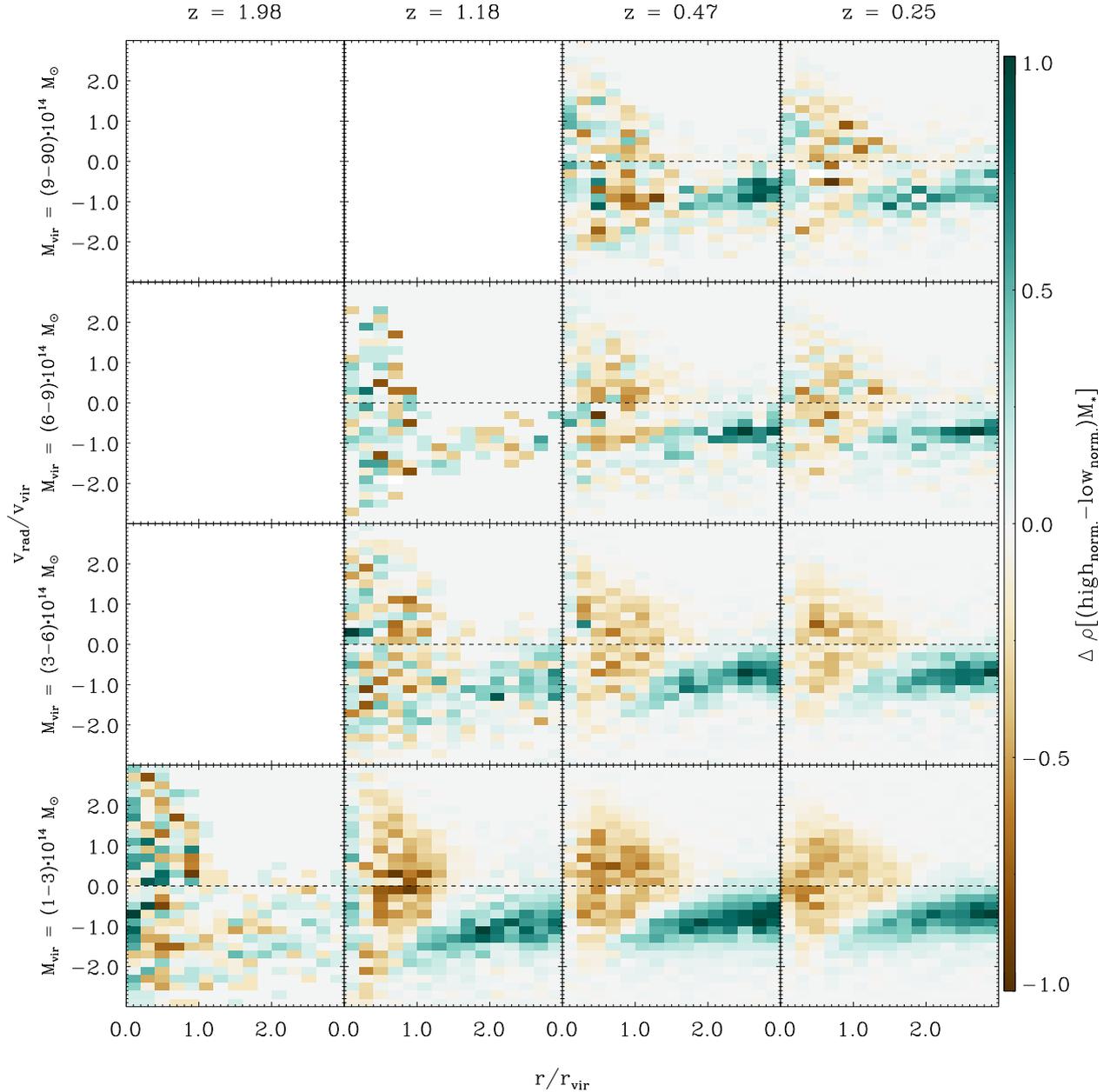}
    \caption{Combined Box2/hr and Box2b/hr comparison of high and low stellar mass of quiescent satellite galaxies. Figure \ref{fig:PSGrid_r} was split into a high ($M_{*} > 1.5 \cdot 10^{10} \, \mathrm{M_{\odot}}$) and low ($M_{*} < 1.5 \cdot 10^{10} \, \mathrm{M_{\odot}}$) stellar mass bin. Thereafter, the two mass bins were normalised to the maximum value of each panel and subsequently subtracted from one another, resulting in maps of relative normalised stellar mass densities, i.e. $\Delta \rho [\mathrm{({high}_{norm.}-{low}_{norm.})M_*}]$.}
    \label{fig:PSGrid_r_comparison}
\end{figure*}

\begin{figure*}
	\includegraphics[width=2.0\columnwidth]{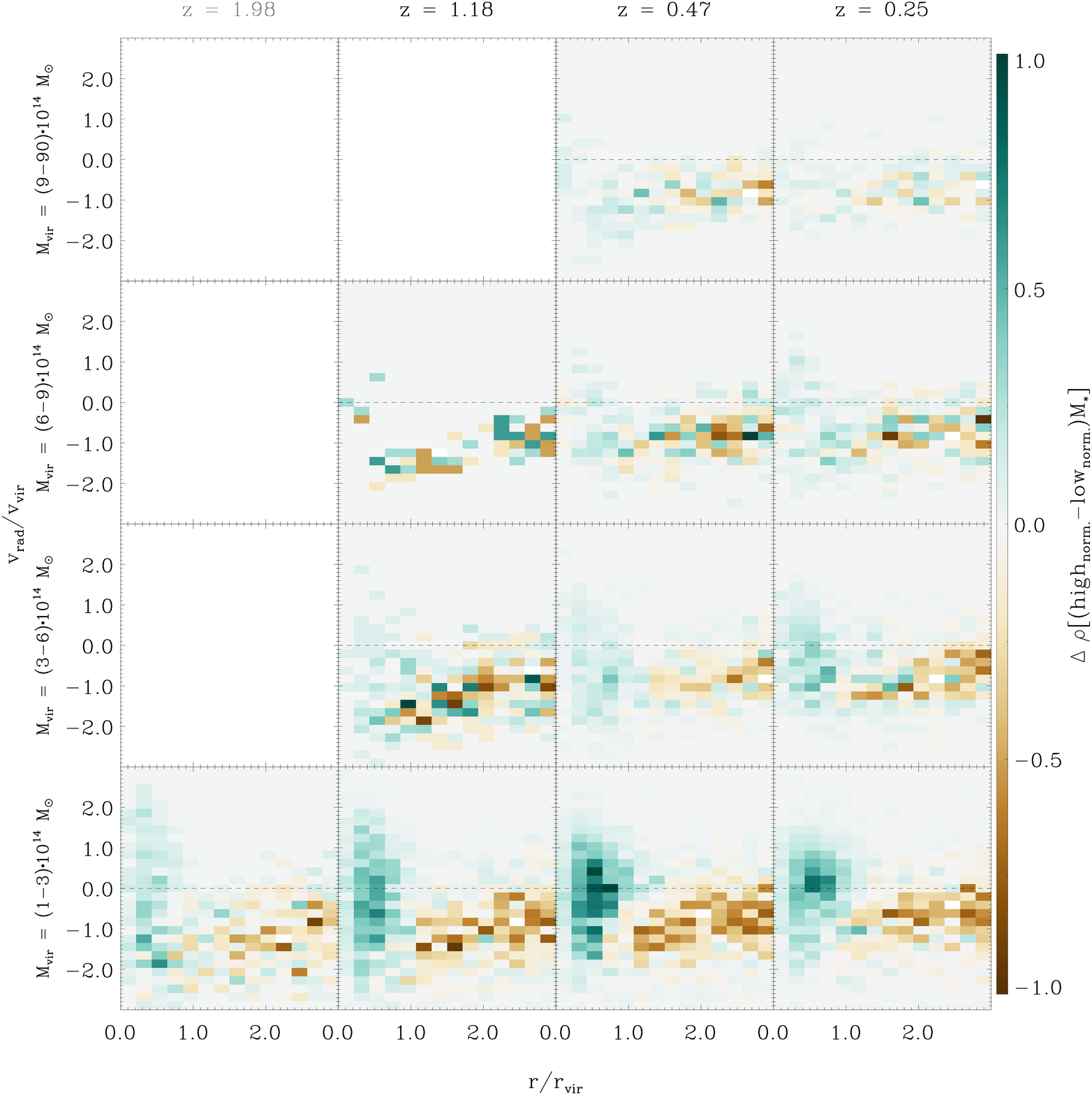}
    \caption{Combined Box2/hr and Box2b/hr comparison of high and low stellar mass of star-forming satellite galaxies. Figure \ref{fig:PSGrid_b} was split into a high ($M_{*} > 1.5 \cdot 10^{10} \, \mathrm{M_{\odot}}$) and low ($M_{*} < 1.5 \cdot 10^{10} \, \mathrm{M_{\odot}}$) stellar mass bin. Thereafter, the two mass bins were normalised to the maximum value of each panel and subsequently subtracted from one another, resulting in maps of relative normalised stellar mass densities, i.e. $\Delta \rho [\mathrm{({high}_{norm.}-{low}_{norm.})M_*}]$.}
    \label{fig:PSGrid_b_comparison}
\end{figure*}

In Figures \ref{fig:PSGrid_r_comparison} and \ref{fig:PSGrid_b_comparison}, we investigate the effect of splitting the radial phase space of satellite galaxies into high and low stellar mass samples ($M_{*} > 1.5 \cdot 10^{10} \, \mathrm{M_{\odot}}$ and $M_{*} < 1.5 \cdot 10^{10} \, \mathrm{M_{\odot}}$).
As shown in Figure \ref{fig:PSGrid_r_comparison}, we find a relative excess of high stellar mass quiescent satellites outside the virial radius, as indicated by the cyan colour. In contrast, the quiescent population within the clusters is characterised by an excess of low stellar mass satellites.

The excess of low stellar mass quiescent galaxies below the virial radius is likely driven by the dynamical friction timescale. More massive satellite galaxies merge with the BCG on much faster timescales than lower mass satellite galaxies.
This is the result of the dynamical friction coefficient being of order of the reciprocal of the time of relaxation of a system, i.e. the dynamical friction timescale is ${t_{\mathrm{dyn fric}} \sim 1/M}$, where $M$ is the sum of the masses of the two objects under consideration \citep{1943ApJ....97..255C}.
Consequently, fewer high mass quiescent galaxies are found below the virial radius in Figure \ref{fig:PSGrid_r_comparison}.
For a discussion regarding the effect of dynamical friction on orbital decay times see Section \ref{sub:track_survivors}.

The apparent lack of high stellar mass quiescent satellites below the virial radius is also driven by the fact that satellite galaxies within clusters no longer experience mass accretion. Instead, after losing their dark matter halo, satellite galaxies in clusters often experience stellar mass loss via tidal stripping \citep{2017MNRAS.471.4170T}.
As a result, the inner cluster population is biased towards lower stellar mass galaxies.

In contrast to Figure \ref{fig:PSGrid_r_comparison}, Figure~\ref{fig:PSGrid_b_comparison} shows a high stellar mass excess within the cluster, while the outskirts are characterised by a relative excess of low stellar mass star-forming satellites.
The overabundance of high stellar mass star-forming satellites within clusters, especially at lower cluster masses, is driven by a selection effect: a higher stellar mass implies a stronger gravitational binding energy and, thus, protects against ram-pressure stripping more effectively. Ultimately, low stellar mass satellite galaxies are less likely to remain star-forming after in-fall and, hence, are less likely to be found within clusters (see Section~\ref{sub:track_survivors}).
This effect becomes less significant in higher mass clusters as ram-pressure stripping becomes more violent, and stellar mass shielding less effective (see Section~\ref{sec:track}).

We verified that the excess of low stellar mass quiescent satellite galaxies in the inner regions of clusters is not solely driven by a combination of the normalisation and in-falling low stellar mass galaxies being preferentially star-forming. The individual panel normalisation, which implies that any local excess must be compensated by a local deficit within a given panel, contributes to a stronger visual impact in some panels. However, it is not the dominant driver behind the behaviour found in Figures \ref{fig:PSGrid_r_comparison} and \ref{fig:PSGrid_b_comparison}.

\section{Ram-pressure stripping and satellite galaxy quenching prior to in-fall}
\label{sec:track}

\subsection{Tracking in-falling star-forming satellite galaxies}

\begin{figure}
	\centering
	\includegraphics[width=\columnwidth]{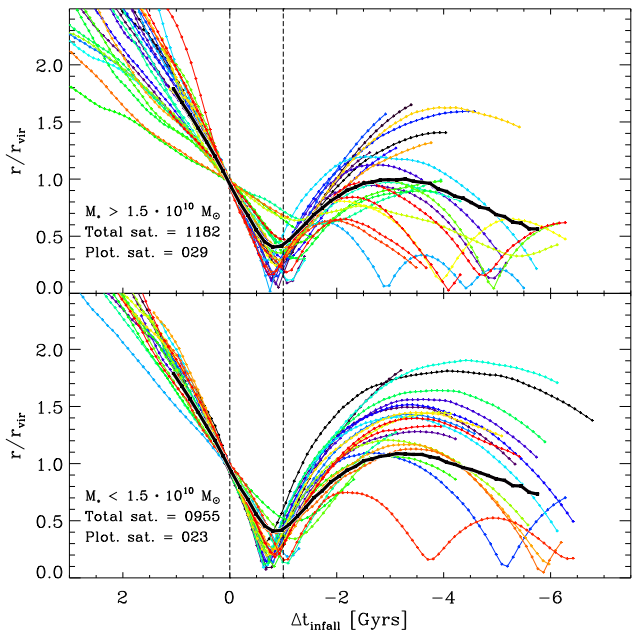}
    \caption{Cluster-centric 3D radial distance scaled to the temporal evolution, i.e. deviation to in-fall time $\Delta \mathrm{t_{infall}}$, of high and low mass satellite galaxies extracted from Box2/hr. Each line represents an individual satellite galaxy tracked through time from $z = 1.01$ to the present day. Only a selection of satellite galaxy trajectories are plotted for facilitated visualisation. The satellite galaxies are scaled to the point where they pass below $1 \, \mathrm{R_{vir}}$ and this time is set to $\Delta \mathrm{t_{infall}} = 0$. Top panel: satellite galaxies with stellar mass $M_* > 1.5 \cdot 10^{10} \, \mathrm{M_{\odot}}$. Bottom panel: satellite galaxies with stellar mass $M_* < 1.5 \cdot 10^{10} \, \mathrm{M_{\odot}}$. The thick black line depicts the mean value of the entire satellite galaxy population in the given mass range, including the trajectories not plotted. The dashed vertical lines indicate the points in time corresponding to in-fall and $1 \, \mathrm{Gyr}$ after in-fall.}
    \label{fig:OrbitTrack}
\end{figure}

\begin{figure}
	\includegraphics[width=\columnwidth]{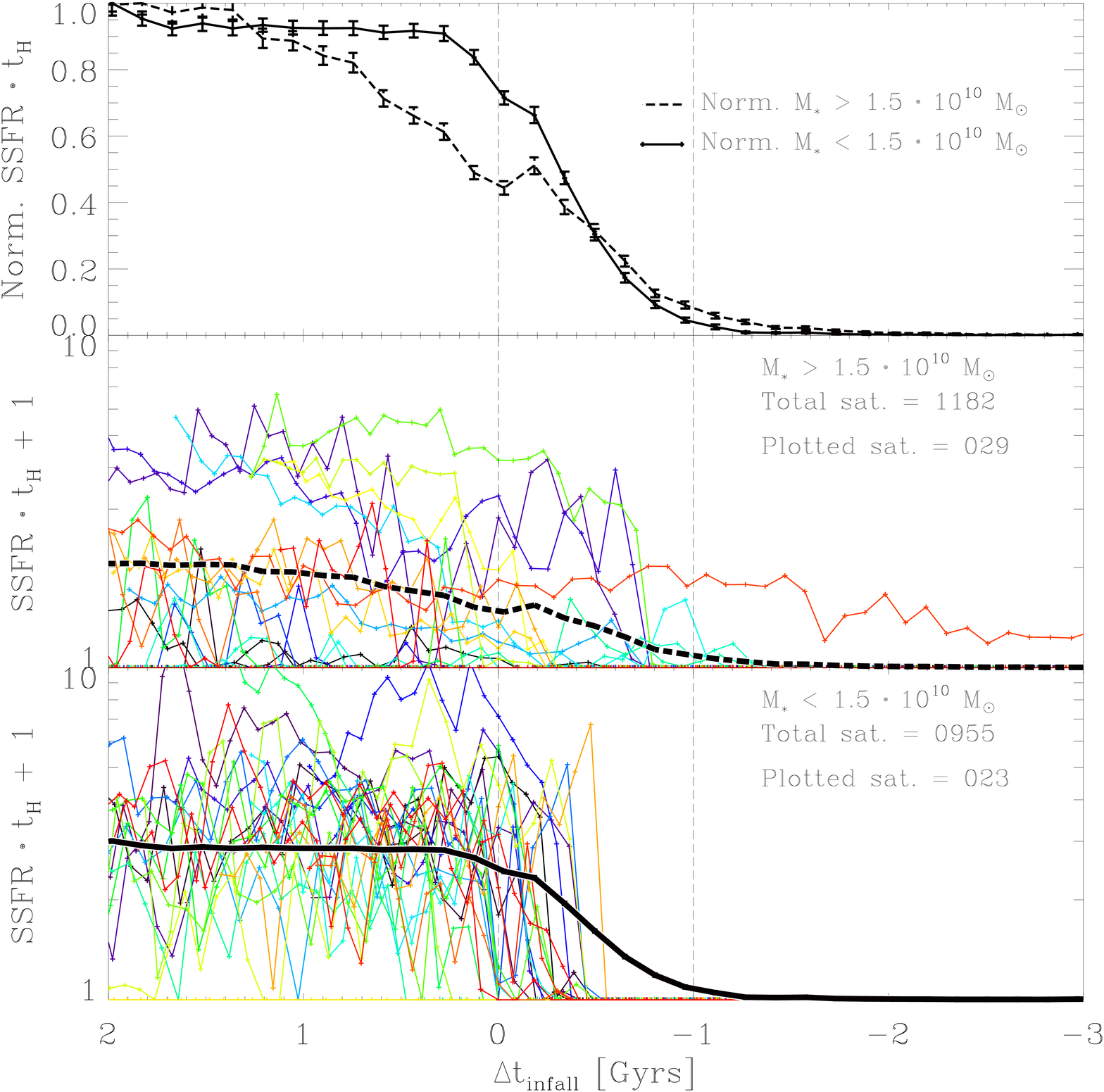}
    \caption{Blueness scaled to the temporal evolution, i.e. deviation to in-fall time $\Delta \mathrm{t_{infall}}$, of high and low mass satellite galaxies extracted from Box2/hr. Each line represents an individual satellite galaxy tracked through time from $z = 1.01$ to the present day. Only a selection of satellite galaxy trajectories are plotted for facilitated visualisation. The satellite galaxies are scaled to the point where they pass below $1 \, \mathrm{R_{vir}}$ and this time is set to $\Delta \mathrm{t_{infall}} = 0$. The thick black lines depict the mean value of the entire satellite galaxy population in the given mass range, including the trajectories not plotted. Top panel: the two black lines were normalised to their maximum value, as such they are linearly scaled versions of the two thick black lines in the middle and bottom panel. The error bars were obtained through bootstrapping. Middle panel: satellite galaxies with stellar mass $M_* > 1.5 \cdot 10^{10} \, \mathrm{M_{\odot}}$. Bottom panel: satellite galaxies with stellar mass $M_* < 1.5 \cdot 10^{10} \, \mathrm{M_{\odot}}$. The dashed vertical lines indicate the points in time corresponding to in-fall and $1 \, \mathrm{Gyr}$ after in-fall.}
    \label{fig:OrbitTrackBlue}
\end{figure}

To better understand the nature of the quenching mechanisms at play, we track individual satellite orbits from the outskirts of clusters to long after they have passed the virial radius. Figure \ref{fig:OrbitTrack} shows the radial distance evolution of two different mass populations of galaxies as they fall into their respective clusters. Figure \ref{fig:OrbitTrackBlue} follows the blueness evolution of the same galaxies. 

Both figures scale the individual satellite orbits to the point where they cross the virial radius. Specifically, the point in time is identified where the individual galaxies cross the cluster virial radius and the trajectory of the now satellite galaxies is then scaled so that all galaxies cross the virial radius at a deviation to in-fall time of $\Delta \mathrm{t_{infall}} = 0$. 

The galaxies are selected according to three criteria: a.) host cluster mass at redshift $z = 1.01$ has to be in the range of ${1 \cdot 10^{14} < M_{\mathrm{vir}}/\mathrm{M_{\odot}} < 3 \cdot 10^{14}}$; b.) the satellite galaxies must be located at a radial interval of ${1.5 < r/r_{\mathrm{vir}} < 4.5}$ at ${z=1.01}$ before descending into the cluster; c.) satellite galaxies must have at least 100 stellar particles (${M_{*} > 3.5 \cdot 10^{9} \, h^{-1} \mathrm{M_{\odot}}}$) and be considered star-forming, i.e. $\mathrm{SSFR} \cdot t_\mathrm{H} > 0.3$.

In order to evaluate the quenching efficiency of different galaxy masses, the tracked satellite galaxies are further subdivided into two populations. The high stellar mass population is characterised by satellite galaxies which consistently, i.e. independent of delta in-fall time, have a stellar mass above $M_{*} > 1.5 \cdot 10^{10} \, \mathrm{M_{\odot}}$, while, the low mass population in Figures \ref{fig:OrbitTrack} and \ref{fig:OrbitTrackBlue} consistently has a stellar mass of $M_{*} < 1.5 \cdot 10^{10} \, \mathrm{M_{\odot}}$. It is important to note that if a galaxy crosses the stellar mass threshold $M_{*} = 1.5 \cdot 10^{10} \, \mathrm{M_{\odot}}$, it is no longer considered a member of either group. This is done so as to filter out galaxies that experience significant mass growth and, thus, are likely prone to quenching as a result of internal changes, rather than through the cluster environment, but also filters out satellite galaxies that experience mass loss through tidal stripping.

Figure \ref{fig:OrbitTrack} reveals no differences in the behaviour between the high (upper panel) and low (lower panel) stellar mass satellite galaxy populations. Satellite galaxies are accreted onto the cluster and, depending on their initial in-fall conditions, exhibit different orbital behaviour.
The stellar mass, thus, likely has no impact on the radial orbital evolution. This is best exemplified by the wide range of orbital periods in both populations, ranging from shallow tangential orbits, with long orbital time-scales and a relatively constant large radial distance, to very radial orbits with short orbital periods that rapidly decrease in amplitude.

However, if we look at the blueness of satellite galaxies (Figure \ref{fig:OrbitTrackBlue}) it becomes apparent that low stellar mass satellite galaxies generally have a higher average specific star-formation compared to their high mass counterparts prior to in-fall.
At lower stellar mass, i.e. lower stellar particle numbers, increasing resolution effects become non-negligible.
As a result, low stellar mass satellite galaxies experience a more bursty SFR than high mass satellite galaxies, for which a more continuous SSFR can be seen.
After in-fall into the cluster's virial radius all satellite galaxies show a strong decline in SSFR, with most galaxies being quenched completely $1 \, \mathrm{Gyr}$ after in-fall. 

The thick black lines in the middle (dashed line) and bottom (solid line) panel in Figure \ref{fig:OrbitTrackBlue} describe the average behaviour of each population in response to quenching. 
We confirmed that the mean SSFR provides a good proxy for individual galaxy behaviour.
To stronger emphasise the difference between the high and low stellar mass populations, the thick black lines are normalised to their respective maximum value and shown in the top panel. 

The high stellar mass satellite galaxies experience a far earlier onset of quenching, characterised by a continuous steady decline in star-formation, rather than a rapid decrease. In contrast, the low stellar mass satellite galaxies are characterised by a constant star-formation until shortly before falling into the cluster ($\Delta \mathrm{t_{infall}} \sim 0.3$), only to experience rapid quenching with in-fall. 
As the termination of star-formation is effectively instantaneous for many individual low stellar mass galaxies (coloured lines in bottom panel of Figure \ref{fig:OrbitTrackBlue}), an extended quenching timescale for individual low mass galaxies is not available. Therefore, the average, as shown in the bottom panel of Figure \ref{fig:OrbitTrackBlue}, reflects the average decline of the entire population rather than individual quenching timescales of low stellar mass galaxies. However, higher resolution simulations likely will extend the quenching timescale of low stellar mass satellite galaxies.

Independent of the stellar mass, almost all satellite galaxies in Figure \ref{fig:OrbitTrackBlue} are quenched within $1 \, \mathrm{Gyr}$ after in-fall. As a population rather than individually, low stellar mass satellite galaxies are quenched on a time-scale of $t_{\mathrm{low}} \sim 1 \, \mathrm{Gyr}$, while the high stellar mass satellite galaxies are quenched on a time-scale roughly twice as long: $t_{\mathrm{high}} \sim 2-3 \, \mathrm{Gyr}$, due to the earlier onset of their quenching. This shows that stellar mass is relevant with regard to the quenching mechanisms at play, as well as the time-scales on which they occur.
This is discussed further in Section \ref{sub:categories}, where the impact of mass quenching and environmental quenching is investigated.

Once commenced, the decrease in specific star-formation for both mass bins appears to be linear. \cite{2010MNRAS.406.2249W} find that the diffuse gas in their model must be stripped linearly, rather than exponentially, in order to reproduce observations. This implies that stripping becomes more efficient the longer a galaxy has been a satellite \citep{2010MNRAS.406.2249W}.
This is in agreement with findings from \cite{2018MNRAS.475.3654Z}, which state that, due to the gas depletion times, it is quite plausible for galaxies crossing the cluster's virial radius during in-fall to be quenched as a result of halo gas removal. Subsequently, this also explains the abundance of quenched galaxies, especially disks, in the outskirts of clusters \citep{2018MNRAS.475.3654Z}. 
\cite{2012MNRAS.424..232W} and \cite{2013MNRAS.432..336W} find that satellites remain active after in-fall on time-scales longer than the quenching time-scale. Once the satellite SFR fading process has begun, it is rapid and likely shorter than $\sim 2.4 \, \mathrm{Gyr}$ \citep{2012MNRAS.424..232W}. In contrast to our findings this `delayed-then-rapid' quenching scenario finds that satellites evolve unaffected for $2-4 \, \mathrm{Gyr}$ after in-fall before being rapidly quenched \citep{2013MNRAS.432..336W}.

The gradual decline in specific star-formation of the high stellar mass satellite galaxies in Figure \ref{fig:OrbitTrackBlue} is characterised by a statistical anomaly. Namely, a bump in specific star-formation is visible around $0.2 \, \mathrm{Gyr}$ after crossing the virial radius. This is in agreement with numerical studies that have found evidence for an enhancement of star-formation as a result of the additional pressure exerted by the intra-cluster-medium (ICM) on the inter-stellar-medium (ISM) \citep{2003ApJ...596L..13B, 2008A&A...481..337K, 2008MNRAS.389.1405K, 2009A&A...499...87K,2009ApJ...694..789T}, however \cite{2012MNRAS.422.1609T} find no evidence for a burst in star-formation. 

The short time-scale quenching mechanism responsible for both the star-burst and the inner cluster region rapid shutdown in star-formation is ram-pressure stripping.
Originally defined by \cite{1972ApJ...176....1G}, ram-pressure stripping is the mechanism whereby the cold gas component is removed from the galaxy as a result of an increased pressure gradient, $P_{\mathrm{ram}}$, exerted onto the galaxy by the ICM:
\begin{ceqn}
\begin{align}
\label{eq:ram}
P_{\mathrm{ram}} \approx \rho_{\mathrm{ICM}} \Delta v^2
\end{align}
\end{ceqn}

\noindent
where $\rho_{\mathrm{ICM}}$ is the ICM density and $\Delta v^2$ is the three dimensional galaxy velocity with respect to the cluster centre. 

Generally, a typical spiral will be able to retain material within its plane if the force per unit area, i.e. the pressure $P_{retain}$, does not exceed \citep{1972ApJ...176....1G}:
\begin{ceqn}
\begin{align}
\label{eq:retain}
P_{\mathrm{retain}} = 2 \pi G \sigma_{\mathrm{star}}(r) \sigma_{\mathrm{gas}}(r)
\end{align}
\end{ceqn}

\noindent
where $\sigma_{\mathrm{star}}(r)$ is the star surface density at a given radius and $\sigma_{\mathrm{gas}}(r)$ is the gas surface density on the disk at a given radius. Hence, ram-pressure stripping occurs when $P_{\mathrm{ram}} > P_{\mathrm{retain}}$. It also follows that ram-pressure becomes more efficient when a galaxy exposes more of its surface area to $P_{\mathrm{ram}}$.

The star-burst seen in the high stellar mass, and to a smaller extent in the low stellar mass, satellite galaxies in Figure \ref{fig:OrbitTrackBlue} is probably driven by the onset of ram-pressure stripping. Low stellar mass galaxies cannot shield themselves as effectively as high stellar mass galaxies from ram-pressure. As a result high stellar mass galaxies can retain more gas at the same ram-pressure. Hence, the onset of ram-pressure stripping more likely leads to compression than to expulsion in high stellar mass galaxies; whereas in low stellar mass galaxies the likelihood of quickly losing the cold gas component is higher. To strengthen these assumptions, we study an independent population of satellite galaxies, selected as the discussed sample but in a cluster mass range of $3 \cdot 10^{14} < M_{\mathrm{vir}}/\mathrm{M_{\odot}} < 6 \cdot 10^{14}$ at $z=1.01$. We find the same behaviour, i.e. this independent satellite galaxy population is also characterised by a star-burst at $\Delta \mathrm{t_{infall}} = -0.2$, strongly suggesting a physical origin.
Furthermore, we could not find any correlation between pericentre passage and a star-burst (see Figure \ref{fig:desc_peri}). For more details concerning the onset of ram-pressure stripping and the subsequent induced star-burst in cluster satellite galaxies see our second paper Lotz et al. in prep.

\subsection{Tracking survivors}
\label{sub:track_survivors}

\begin{figure}.
	\includegraphics[width=\columnwidth]{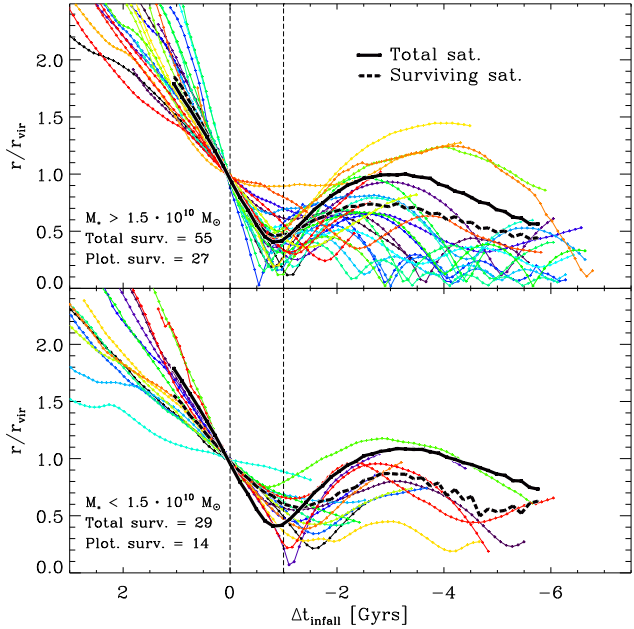}
    \caption{Surviving sample's cluster-centric 3D radial distance scaled to the temporal evolution, i.e. deviation to in-fall time $\Delta \mathrm{t_{infall}}$, of high and low mass satellite galaxies extracted from Box2/hr. Each line represents an individual surviving satellite galaxy tracked through time from $z = 1.01$ to the present day. All surviving satellite galaxy trajectories, i.e. all satellite galaxies that are still considered star-forming $1 \, \mathrm{Gyr}$ after in-fall are plotted. The satellite galaxies are scaled to the point where they pass below $1 \, \mathrm{R_{vir}}$ and this time is set to $\Delta \mathrm{t_{infall}} = 0$. Top panel: surviving satellite galaxies with stellar mass $M_* > 1.5 \cdot 10^{10} \, \mathrm{M_{\odot}}$. Bottom panel: surviving satellite galaxies with stellar mass $M_* < 1.5 \cdot 10^{10} \, \mathrm{M_{\odot}}$. The dashed vertical lines indicate the points in time corresponding to in-fall and $1 \, \mathrm{Gyr}$ after in-fall.}
    \label{fig:SurvTrack}
\end{figure}

\begin{figure}
	\includegraphics[width=\columnwidth]{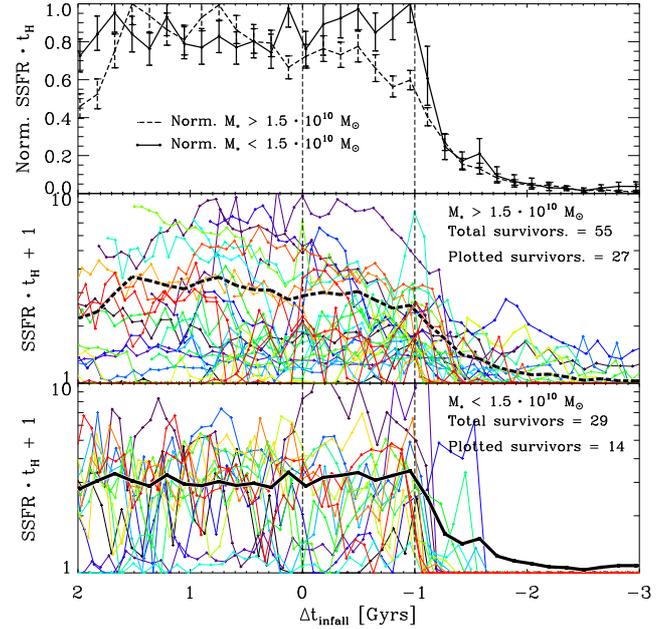}
    \caption{Surviving sample's blueness scaled to the temporal evolution, i.e. deviation to in-fall time $\Delta \mathrm{t_{infall}}$, of high and low mass satellite galaxies extracted from Box2/hr. Each line represents an individual surviving satellite galaxy tracked through time from $z = 1.01$ to the present day. All surviving satellite galaxy trajectories, i.e. all satellite galaxies that are still considered star-forming $1 \, \mathrm{Gyr}$ after in-fall are plotted. The satellite galaxies are scaled to the point where they pass below $1 \, \mathrm{R_{vir}}$ and this time is set to $\Delta \mathrm{t_{infall}} = 0$. The thick black lines depict the mean value of the surviving satellite galaxy population in the given mass range. Top panel: the two black lines were normalised to their maximum value, as such they are linearly scaled versions of the two thick black lines in the middle and bottom panel. The error bars were obtained through bootstrapping. Middle panel: surviving satellite galaxies with stellar mass $M_* > 1.5 \cdot 10^{10} \, \mathrm{M_{\odot}}$. Bottom panel: surviving satellite galaxies with stellar mass $M_* < 1.5 \cdot 10^{10} \, \mathrm{M_{\odot}}$. The dashed vertical lines indicate the points in time corresponding to in-fall and $1 \, \mathrm{Gyr}$ after in-fall.}
    \label{fig:SurvTrackBlue}
\end{figure}

To better understand the quenching mechanisms at work, we introduce a criterion to identify satellite galaxies that survive, i.e. remain unquenched, much longer than the average of the total population. The surviving satellite population is defined as galaxies that are still considered star-forming $1 \, \mathrm{Gyr}$ after in-fall, as indicated by the dashed vertical lines at $\Delta \mathrm{t_{infall}} = -1$ in Figures \ref{fig:SurvTrack} and \ref{fig:SurvTrackBlue}. Figures \ref{fig:SurvTrack} and \ref{fig:SurvTrackBlue} study the distinguishing attributes of the surviving satellite galaxies, in comparison to the previously discussed total population in Figures \ref{fig:OrbitTrack} and \ref{fig:OrbitTrackBlue}. This allows the study of quenching inhibitors, resulting in a better understanding of the physical quantities responsible for prolonged survival in a galaxy cluster environment.

We find that high stellar mass satellite galaxies (${M_* > 1.5 \cdot 10^{10}\, \mathrm{M_{\odot}}}$) preferentially survive if they have a very high stellar mass, compared to the mean stellar mass of the entire high stellar mass population (see Table \ref{tab:stmass}). This implies that stellar mass correlates strongly with the ability to remain star-forming, i.e. survival during in-fall. This is reflected by the significant mass difference (${\mathrm{factor} \sim 4}$) between the high stellar mass survivors and the total high stellar mass population.
This suggests that a minimum stellar mass threshold needs to be reached to be shielded from ram-pressure stripping. 

For low mass satellite galaxies, we find no meaningful correlation between continued star-formation and stellar mass. However, low stellar mass satellite galaxies ($M_* < 1.5 \cdot 10^{10}\, \mathrm{M_{\odot}}$) show a correlation between the in-fall orbits and their subsequent survival. As the contrast between solid and dashed black lines in Figure \ref{fig:SurvTrack} indicates, surviving low stellar mass satellite galaxies are likely to have shallow orbits, i.e. small radial distance fluctuations and larger pericentres, while their high stellar mass counterparts do not show this behaviour strongly within the first Gyr after in-fall. However, the comparison between the total (solid line) and surviving (dashed line) high stellar mass populations in the top panel in Figure \ref{fig:SurvTrack} shows that dynamical friction is more effective for higher mass objects. 
Within the first Gyr after in-fall the two populations in the top panel are characterised by very similar orbits and pericentres. After the satellite galaxies have been part of the cluster for over a Gyr, their orbits show divergence.
As the high stellar mass survivors are on average $\sim 4$ times more massive than the average total high stellar mass population, they experience dynamical friction to a stronger degree (see Section \ref{sub:PS_stellar}). Subsequently, the orbits of the high stellar mass survivors are circularised more effectively.

In summary, we conclude that above a certain stellar mass threshold satellite galaxy survival correlates strongly with stellar mass, whereas below this threshold survival is far more dependent on the in-fall orbit. This implies that stellar mass and shallow orbits shield satellite galaxies from ram-pressure stripping with different efficiencies depending on the satellite galaxies in question. Furthermore, it suggests that stellar mass is more efficient in shielding than the orbital configuration. This is supported by the fact that fewer low stellar mass satellite galaxies survive relative to the total low stellar mass population than is the case for the high stellar mass comparison. Specifically, $4.7$ per cent of the high stellar mass population survive, whereas only $3.0$ per cent of the low stellar mass satellite galaxies survive up to $1 \, \mathrm{Gyr}$ after in-fall.

Considering equations \ref{eq:ram} and \ref{eq:retain}, it is also clear that more massive galaxies (with higher $M_*$) have larger radii, $r$, at which the ram-pressure, $P_{\mathrm{ram}}$, is equivalent to $P_{\mathrm{retain}}(r)$. This simple theoretical description implies that high stellar mass galaxies are more efficient at shielding the cold gas component from ram-pressure stripping, as $P_{\mathrm{retain}}(r)$ is larger for a given radius for more massive galaxies. As such, Magneticum is in line with simple theoretical predictions. 

In contrast to the gradual decline found for the total massive satellite galaxy population, no gradual decline is observed in either population in Figure \ref{fig:SurvTrackBlue}. The survival criterion likely filters out any satellite galaxies that experienced quenching in the outskirts of the cluster, as this would inhibit the survival at in-fall.

\begin{table}
	\centering
	\begin{tabular}{lll}
		\hline
		Sample $[\mathrm{M_{\odot}}]$ & Mean st. mass $[\mathrm{M_{\odot}}]$ & Numb. of sats. \\
		\hline
    $M_* > 1.5 \cdot 10^{10}$ All   & $6.01 \cdot 10^{10}$ &   1182\\
    $M_* > 1.5 \cdot 10^{10}$ Surv. & $2.38 \cdot 10^{11}$ &   55  \\
    $M_* < 1.5 \cdot 10^{10}$ All   & $8.58 \cdot 10^9$    &   955 \\
    $M_* < 1.5 \cdot 10^{10}$ Surv. & $9.14 \cdot 10^{9}$  &   29  \\
		\hline
	\end{tabular}
    \caption{Table listing the relevant information of the samples depicted in Figures \ref{fig:OrbitTrack}, \ref{fig:OrbitTrackBlue}, \ref{fig:SurvTrack} and \ref{fig:SurvTrackBlue}. The first column references the sample, i.e. the stellar mass range and which population is under consideration. The second column displays the mean stellar satellite galaxy mass of the sample selected in the first column. The third column lists the number of satellite galaxies in a given selected sample. Each satellite galaxy has a stellar mass of at least $M_* > 3.5 \cdot 10^{9} \, h^{-1} \mathrm{M_{\odot}}$.}
     \label{tab:stmass}
\end{table}

\begin{figure}
    \centering
	\includegraphics[width=0.99\columnwidth]{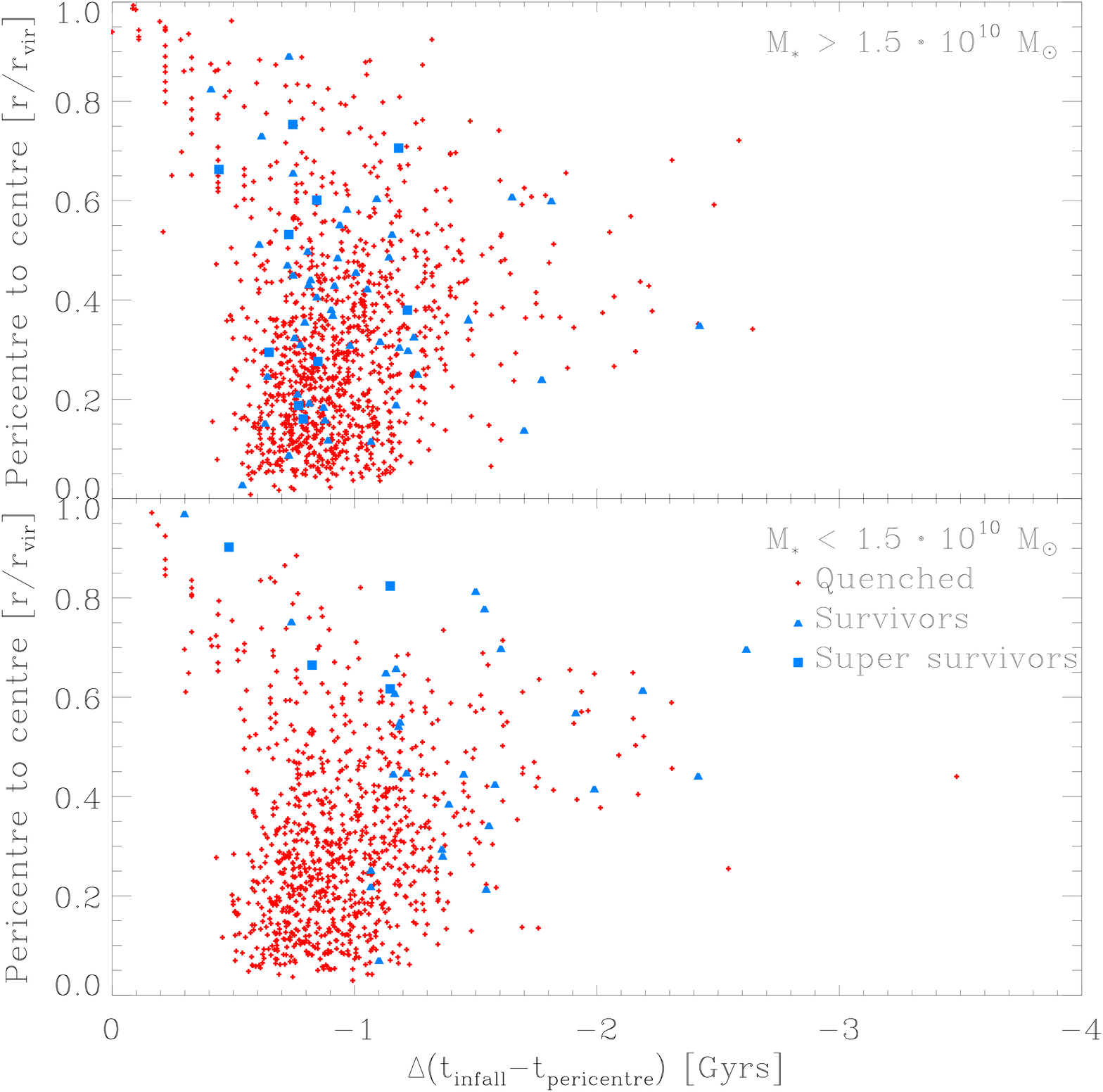}
    \caption{Satellite galaxy pericentre to centre of the cluster distance as a function of difference in time between in-fall and pericentre. Red crosses indicate all satellite galaxies which are quenched within $1\,$Gyr after in-fall. Blue triangles indicate all satellite galaxies which remain star-forming for longer than $1\,$Gyr after in-fall, while blue squares indicate satellite galaxies which remain star-forming longer than $1\,$Gyr after passing their pericentre. Top bin shows high stellar mass ($M_* > 1.5 \cdot 10^{10} \mathrm{M_{\odot}}$) and bottom bin shows low stellar mass ($M_* < 1.5 \cdot 10^{10} \mathrm{M_{\odot}}$) satellite galaxies.}
    \label{fig:desc_scatter}
\end{figure}

In Figure \ref{fig:desc_scatter}, we study the impact of galactic orbits on survival. Specifically, Figure \ref{fig:desc_scatter} shows the cluster's virial radius normalised distance ($r/r_{\mathrm{vir}}$) between an individual satellite galaxy's pericentre and the cluster centre as a function of the difference in time between in-fall and pericentre passage, $\Delta \mathrm{(t_{infall}-t_{pericentre})}$.
Similarly to previous plots, the sample is divided into the standard high and low stellar mass bins. In addition to the previously defined 'survivors', we define another category, namely, the 'super survivors'. Super survivors are defined as the subsample of satellite galaxies that are still star-forming $1\,$Gyr after pericentre passage.

Figure \ref{fig:desc_scatter} shows that orbital characteristic are more important in the low stellar mass bin than they are in the high stellar mass bin. In the high stellar mass bin there we find no evidence for a significant correlation between pericentre height or $\Delta \mathrm{(t_{infall}-t_{pericentre})}$ and survival (blue triangles). However, super survivors (blue squares) show a preference for high pericentres. The low stellar mass survivors are characterised by a strong preference towards high pericentres and high $\Delta \mathrm{(t_{infall}-t_{pericentre})}$, supporting our previous findings that orbits are far more relevant in the low stellar mass regime. Low stellar mass super survivors all have pericentres above $0.6 \, \mathrm{r/r_{vir}}$, indicating a strong preference towards shallow orbits. To summarise, satellite galaxy orbits are especially important in the low stellar mass regime, as stellar mass shielding against ram-pressure stripping, which is especially violent near the cluster centre, becomes less efficient.

\subsection{Impact of in-fall environment}
\label{sub:categories}

\begin{figure}
    \centering
	\includegraphics[width=0.99\columnwidth]{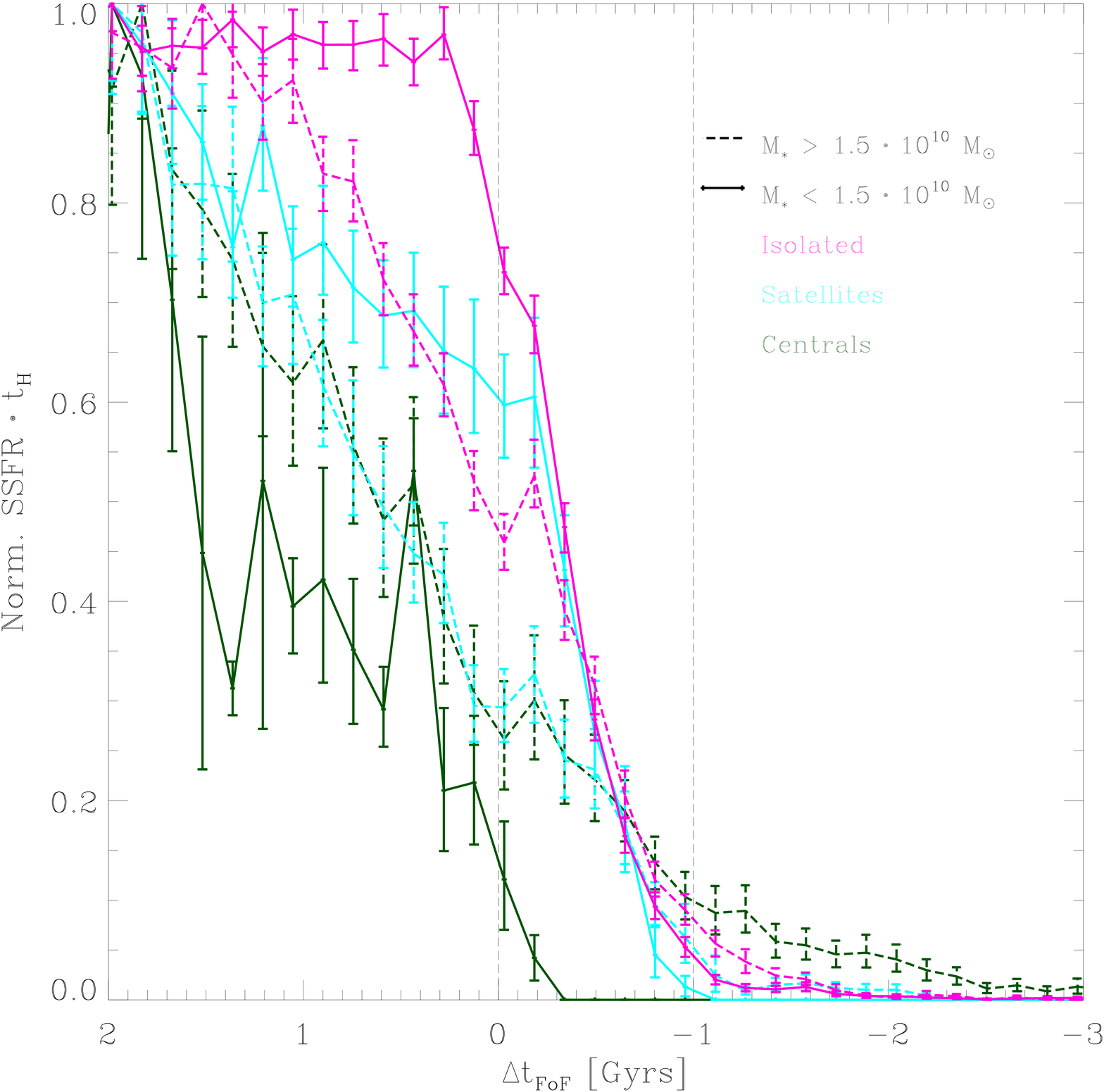}
    \caption{Same as Figure \ref{fig:OrbitTrackBlue} but displaying different in-fall categories of satellite galaxies normalised to the point where \texttt{SUBFIND} identifies FoF membership, i.e. $\Delta \mathrm{t_{FoF}} = 0$. In-falling satellite galaxies are subdivided into three high (dashed line) and low (solid line) stellar mass groups: isolated galaxies (magenta), group satellite galaxies (cyan) and group centrals (green).}
    \label{fig:desc_categories}
\end{figure}

We subdivided galaxies according to their environment at in-fall into three categories: isolated galaxies, group satellite galaxies and group centrals. This categorisation is useful in determining how the in-fall environment impacts the evolution of $\mathrm{SSFR} \cdot t_{\mathrm{H}}$. Similar to Figure \ref{fig:OrbitTrackBlue}, Figure \ref{fig:desc_categories} shows the normalisation to the maximum value of $\mathrm{SSFR} \cdot t_{\mathrm{H}}$ for each case. However, in contrast to previous figures, the in-fall of galaxies has been normalised to the point where \texttt{SUBFIND} identifies galaxies as members of the respective cluster Friends-of-Friends (FoF) halo. Therefore, in-fall into the cluster FoF halo is given by $\Delta \mathrm{t_{FoF}} = 0$. In this section, galaxies are identified via cluster FoF membership rather than virial radius crossing because this facilitates the identification of the environment prior to in-fall. The same population as in Section \ref{sec:track} was used, albeit only a subsample fulfils our group categorisation criterion, as we only consider groups with at least three different galaxies above our standard stellar mass threshold.

As demonstrated by Figure \ref{fig:desc_categories}, the environment at in-fall has a strong impact on the evolution of star-formation. We find that group centrals (green) and group satellites (cyan) experience stronger quenching prior to cluster in-fall, while isolated (magenta) galaxies show a weaker decline in star-formation prior to cluster in-fall. In other words, higher environmental density correlates with stronger quenching prior to cluster in-fall. 
This supports the picture established by the morphology-density relation that a higher density environment is more effective in quenching galaxies \citep{1980ApJ...236..351D, 2003MNRAS.346..601G}.

Similarly to what was shown in Figure \ref{fig:OrbitTrackBlue}, we also find that low stellar mass galaxies (solid line) are typically less quenched prior to in-fall compared to their high stellar mass (dashed line) counterparts. The exception to this is the low stellar mass central galaxy population (green solid line), which only has a sample size of five, and is characterised by the strongest decline in star-formation prior to in-fall. Excluding the small low stellar mass central galaxy population, this suggests that \textit{mass quenching}, i.e. stellar mass dependent mechanisms which are independent of the environment, play an important role in regulating star-formation in cluster outskirts. This is in agreement with findings from \cite{2010ApJ...721..193P} and \cite{2019arXiv190511008L}, who state that at higher stellar masses quenching is controlled by stellar mass, while at lower stellar masses the halo mass, i.e. environment, is the determining factor in galaxy quenching.
Findings based on a smaller Magneticum box with higher resolution further support this notion, stating that environmental quenching is more important for satellite galaxies than for centrals \citep{2017MNRAS.472.4769T}.

As shown in Figure \ref{fig:OrbitTrack} and \ref{fig:OrbitTrackBlue}, the original sample is split almost evenly between high (1182) and low (955) stellar mass.
When considering the individual sample sizes in Figure \ref{fig:desc_categories}, we find that more dense environments, i.e. groups, are characterised by a relative overabundance of high stellar mass galaxies, while less dense environments, i.e. isolated galaxies, are characterised by a relative overabundance of low stellar mass galaxies. Specifically, we find $91$ high stellar mass group centrals and $167$ high stellar mass group satellites, compared to $5$ low mass centrals and $99$ low mass satellites. In contrast, we find fewer isolated high stellar mass galaxies ($506$) than isolated low stellar mass galaxies ($610$). Essentially, this is a reflection of hierarchical galaxy assembly: more dense environments are more likely to host higher stellar mass galaxies \citep{1997ApJ...490..493N}.

Interestingly, the high stellar mass centrals (green dashed line) remain star-forming the longest after in-fall. This is likely due to two reasons: Firstly, centrals typically are the most massive galaxy within a group, i.e. they have a higher probability of being able to retain their cold gas during ram-pressure stripping (see Section~\ref{sec:track}). Secondly, by definition centrals are embedded within a group environment. The group environment further mitigates the violent effects of the onset of ram-pressure stripping. As such, high stellar mass central galaxies belong to the select group of galaxies which do not experience rapid quenching. This is exemplified by the comparably weak gradient of the dashed green line in Figure \ref{fig:desc_categories}. 

In agreement with our previous results (see Section~\ref{sec:track}), Figure \ref{fig:desc_categories} provides evidence for a star-burst in several populations at $\Delta \mathrm{t_{FoF}} = -0.2$. The star-burst is consistently stronger within the high mass populations. In this picture, a minimum stellar mass is needed to retain the cold gas component for a sufficiently long time to trigger a star-burst, rather than an immediate continuous quenching.

\section{Summary \& discussion}
\label{sec:disc}

\textit{Magneticum Pathfinder} simulations are in fine agreement with a variety of observations, including AGN population properties \citep{2014MNRAS.442.2304H,2015MNRAS.448.1504S} and dynamical properties of galaxies \citep{2015IAUS..309..145R, 2015ApJ...812...29T, 2017MNRAS.464.3742R}. Especially relevant for the scope of this paper are the agreements with \cite{2013A&A...550A.131P, 2014ApJ...794...67M} on the pressure profiles of the intra-cluster medium. Furthermore, both the general shape of the anisotropy profile and its specific values are consistent with the cosmological simulations by \cite{2013MNRAS.429.3079M}.

The analysis of the anisotropy profiles at $z=0.44$ in Figure \ref{fig:BetaBiv} and their development in Figure \ref{fig:BetaGrid} clearly shows that throughout varying redshifts and cluster masses the star-forming population is consistently on more radial orbits than the quiescent population. These findings are in good agreement with \cite{2013A&A...558A...1B}, but stand in opposition to other observations, which find that the quiescent population is more radially dominated than the star-forming population \citep{2017MNRAS.468..364A}. While we find some clusters that behave similarly to the cluster (A85) studied by \cite{2017MNRAS.468..364A}, they are a rare exception. 

The interpretation of the anisotropy behaviour found in Magneticum and \cite{2013A&A...558A...1B} is consolidated by phase space considerations in Section \ref{sec:PS}. 
We not only find good agreement between observations and our simulation (see Figure \ref{fig:ProjPS}), but also find that the star-forming population is overwhelmingly comprised of in-falling satellite galaxies.
The orbits of the star-forming satellite galaxies are radially dominated because, in contrast to their quiescent counterparts, that are less radially dominated, they are experiencing their first passage into the cluster. We find that the vast majority of star-forming satellite galaxies are quenched during their first passage (see Figures \ref{fig:OrbitTrack} and \ref{fig:OrbitTrackBlue}). This is in agreement with observations, which also find that galaxies experience strong quenching during their first passage through the cluster \citep{2011MNRAS.416.2882M, 2013MNRAS.432..336W, 2015MNRAS.448.1715J, 2016MNRAS.461.1202J, 2016MNRAS.463.3083O}.

The global decrease in anisotropy towards smaller radii is likely driven by an orbital selection mechanism.
Of all the galaxies entering a galaxy cluster at a certain time, those that will survive until redshift zero are those characterised by the most tangential orbits at in-fall \citep{2012MNRAS.427.1024I}.
More radially dominated satellite galaxies are likely to experience tidal stripping to a stronger degree, resulting in a decrease in satellite galaxy mass and a subsequent drop below our resolution limit. Thus, the destruction of radially dominated orbits leads to a circularisation of the total population. 
The longer satellite galaxies are members of a given cluster the more likely it is that the more radially dominated subset is selectively destroyed. As older cluster members have a higher likelihood of orbiting at smaller radii, the anisotropy profile decreases towards smaller radii. 
However, due to smaller sample sizes leading to higher bootstrapping errors, the error scatter towards smaller radii increases and, thus, the ability to make meaningful physical deductions is impeded in the innermost regions.

\cite{2016A&A...585A.160A} finds that passive low-mass (${M_{*} < 10^{10} \, \mathrm{M_{\odot}}}$) galaxies at radii $R < 0.3  \, \mathrm{R_{200}}$ are more likely to be on tangential orbits than their high mass ($M_{*} > 10^{10} \, \mathrm{M_{\odot}}$) counterparts. They propose that this may be the result of selective destruction of low mass galaxies on radial orbits through tidal stripping near the cluster centre \citep{2016A&A...585A.160A}. The combination of low mass and small pericentres makes survival near the hostile cluster centre unlikely. As a result, low mass galaxies in the inner regions are only observed on tangential orbits since otherwise they face destruction.
When considering passive satellite galaxies at $z \sim 0.2$ in the same cluster mass range as \cite{2016A&A...585A.160A} ($M_{200} = 7.7^{+4.3}_{-2.7} \cdot 10^{14} \, \mathrm{M_{\odot}}$), we do not replicate the behaviour \citep{2007A&A...467..427P}. 
However, we find a similar behaviour for our star-forming population. Less massive star-forming galaxies are likely to be on more tangential orbits at $r/r_{\mathrm{vir}} \sim 0.5 $ than their more massive counterparts.

The orbital selection effect observed by \cite{2016A&A...585A.160A} for passive galaxies and to a lesser extent by our simulation for star-forming galaxies additionally supports the idea that specific orbits selectively influence the morphological transformation of galaxies. It further explains why red discs are identified by \cite{2017A&A...604A..54K} to inhabit phase space at projected radii between $R_{500}$ and $R_{200}$, a region either associated with recent in-fall or tangential orbits. This would suggest that after blue discs are quenched via ram-pressure stripping during in-fall, thereby becoming red discs, their orbits fundamentally influence their morphological evolution. Provided they are on sufficiently tangential orbits their morphology remains fairly undisturbed. However, if the red disc galaxies are moving on radially dominated orbits they are likely to be disturbed while moving through their pericentres, thus experiencing an accelerated evolution pathway towards elliptical galaxies.

Similarly to \cite{2008MNRAS.383..593M, 2009MNRAS.399.2221B, 2018MNRAS.476.4753J}, we find that ram-pressure stripping efficiencies depend on the orbital parameters and both the satellite and host halo mass. Specifically, high halo masses, low satellite masses and radially dominated orbits ensure more efficient ram-pressure stripping.

In contrast to \cite{2012MNRAS.424..232W} and \cite{2013MNRAS.432..336W}, we find no evidence for `delayed-then-rapid' quenching. Figure \ref{fig:qfrac} suggests that if active, our AGN feedback might be more efficient in quenching compared to observations \citep{2012MNRAS.424..232W}. The combination of stronger AGN feedback and higher quenching efficiencies in Magneticum results in shorter quenching time-scales of satellite galaxies.

The higher quenching efficiency in Magneticum can be the result of three effects associated with the simulation. Firstly, we consider the SSFR rather than the colour of galaxies, resulting in a more immediate representation of star-formation or lack thereof. This results in quenching being measured effectively instantaneously rather than delayed via colour changes. Secondly, the resolution of our simulations do not reproduce morphologies, especially we do not resolve cold thin discs. Consequently, the persistence of cold thin discs against ram-pressure is not properly captured and hence our galaxies are more vulnerable to environmental quenching. Thirdly, stellar mass losses are not modelled to create new gas reservoirs, but rather are distributed to surrounding gas particles. In practice this means that stellar mass loss often is added to already stripped hot gas, further impeding new star-formation.

Similarly to \cite{2017MNRAS.472.4769T}, we find no meaningful signal dependence on whether we sample a spherical or cylindrical volume. We verified this by evaluating the phase space diagrams for both spherical and cylindrical volumes, where the radii of the sphere and cylinder were kept the same ($R_{\mathrm{sph}} = R_{\mathrm{cyl}}$), while the cylinder height $H_{\mathrm{cyl}}$ was varied, as described in the respective sections.

\section{Conclusions}
\label{sec:conc}

In this paper we studied the velocity-anisotropy, the phase space and tracked the orbital behaviour of satellite galaxies in the cluster environment using the hydrodynamical cosmological \textit{Magneticum Pathfinder} simulations. The evaluation of satellite galaxies in clusters of different mass above the mass threshold of $10^{14} \, \mathrm{M_{\odot}}$ at varying redshifts between $z \sim 2$ and present-day has provided a wide statistical sample. Furthermore, satellite galaxy behaviour was compared to observations at $z = 0.44$ and $z = 1.18$ \citep{2013A&A...558A...1B, 2014ApJ...796...65M}. The results can be summarised as follows:
\begin{itemize}

    \item Star-forming satellite galaxies are consistently characterised by more radial orbits than their quiescent counterparts. The velocity-anisotropy profiles show that this behaviour is independent of cluster masses in the studied range $(1-90) \cdot 10^{14} \, \mathrm{M_{\odot}}$ and in the redshift range $2 > z > 0$. Independent of the population under consideration, the velocity-anisotropy profiles are more radially dominated in the outskirts, while tending towards isotropy at smaller radii.

    \item The velocity-anisotropy profile calculated based on the Magneticum simulations is in good agreement with the observations from \cite{2013A&A...558A...1B} at $z=0.44$. Both the dichotomy between star-forming and quiescent satellite galaxies is reproduced, as is the general decrease towards isotropy at smaller radii. The overwhelming majority of simulated data points lie within the $1 \sigma$ confidence regions of the observations.

    \item The line-of-sight phase space comparison shows good agreement between observations and the Magneticum simulations. However, we find a stronger dichotomy between star-forming and quiescent satellite galaxies than the observations conducted by \cite{2013A&A...558A...1B} at $z=0.44$. Nonetheless, the overall behaviour is similar: star-forming satellite galaxies are found predominantly outside $R_{\mathrm{200,crit}}$, while quiescent satellite galaxies overwhelmingly lie within $R_{\mathrm{200,crit}}$.

    \item The radial phase space study in the cluster mass range ${(1-90) \cdot 10^{14} \, \mathrm{M_{\odot}}}$ and in the redshift range $2 > z > 0$ provided an overview of different quenching efficiencies. We find that high cluster mass and, to a lower extent, low redshift increase quenching of the star-forming satellite galaxies. In addition, we find that the vast majority of star-forming satellite galaxies are quenched during their first passage, independent of cluster mass and redshift.

    \item The tracking of individual orbits and their specific star-formation over time demonstrates that satellite galaxies experience strong quenching in the vicinity of the virial radius. The overwhelming majority of satellite galaxies are quenched within $1 \, \mathrm{Gyr}$ after in-fall, i.e. during their first passage. Further, we find that high stellar mass satellite galaxies ($M_* > 1.5 \cdot 10^{10} \, \mathrm{M_{\odot}}$) experience an earlier onset of quenching, leading to a longer quenching time-scale $t_{\mathrm{high}} \sim 2-3 \, \mathrm{Gyr}$ than their low stellar mass counterparts ($M_* < 1.5 \cdot 10^{10} \, \mathrm{M_{\odot}}$), which experience quenching on time-scales $t_{\mathrm{low}} \sim 1 \, \mathrm{Gyr}$.

    \item When solely considering satellite galaxies that `survive', i.e. satellite galaxies that remain star-forming longer than $1 \, \mathrm{Gyr}$ after in-fall, we find a difference between high and low stellar mass satellite galaxies. Low stellar mass satellite galaxies are characterised by atypically shallow orbits. Provided a satellite galaxy is able to maintain a relatively large radial distance $\gtrsim 0.5 \, \mathrm{R_{vir}}$ by having a strongly tangentially supported orbit, it will be quenched on a longer time-scale, i.e. it will be more likely to be strangled than stripped. 

    \item In contrast, high stellar mass satellite galaxies have an increased probability of survival provided they have a very high stellar mass, i.e. they belong to the highest mass subset of the high mass satellite galaxies. Specifically, the $55$ high mass survivors have a mean stellar mass of $M_{*} = 2.38 \cdot 10^{11} \, \mathrm{M_{\odot}}$, while the total high mass population of $1182$ satellite galaxies has a mean stellar mass of $M_{*} = 6.01 \cdot 10^{10} \, \mathrm{M_{\odot}}$, a factor of $\sim 4$ difference.
    
    \item Prior to cluster in-fall, we find that higher environmental density correlates with a stronger decline in star-formation. Specifically, group centrals experience a stronger decline in star-formation than group satellites, which in turn experience stronger quenching than isolated galaxies prior to in-fall. Furthermore, high stellar mass galaxies typically experience stronger quenching prior to in-fall than low stellar mass galaxies, suggesting that mass quenching plays an important role in regulating star-formation in the outskirts of clusters.

\end{itemize}

In summary, the results suggest three fundamental conclusions: Firstly, the dominant quenching mechanism in galaxy clusters is ram-pressure stripping. Secondly, ram-pressure stripping is sufficiently effective to quench the overwhelming majority of star-forming satellite galaxies within $\sim 1 \, \mathrm{Gyr}$ during their first passage. Thirdly, ram-pressure stripping preferentially quenches radial star-forming satellite galaxies.

\section*{Acknowledgements}
We thank Felix Schulze, Tadziu Hoffmann, Ulrich Steinwandel, Matthias Kluge, and the anonymous referee for helpful discussions.
The \textit{Magneticum Pathfinder} simulations were partially performed at the Leibniz-Rechenzentrum with CPU time assigned to the Project `pr86r'. This work was supported by the DFG Cluster of Excellence `Origin and Structure of the Universe', the DFG Cluster of Excellence `From the Origin of the Universe to the First Building Blocks of Life' and by the DAAD, contract number 57396842. We are especially grateful for the support by M. Petkova through the Computational Center for Particle and Astrophysics (C$^2$PAP). Information on the \textit{Magneticum Pathfinder} project is available at \url{http://www.magneticum.org}.



\appendix

\section{ }

\begin{figure}
	\centering
	\includegraphics[width=\columnwidth]{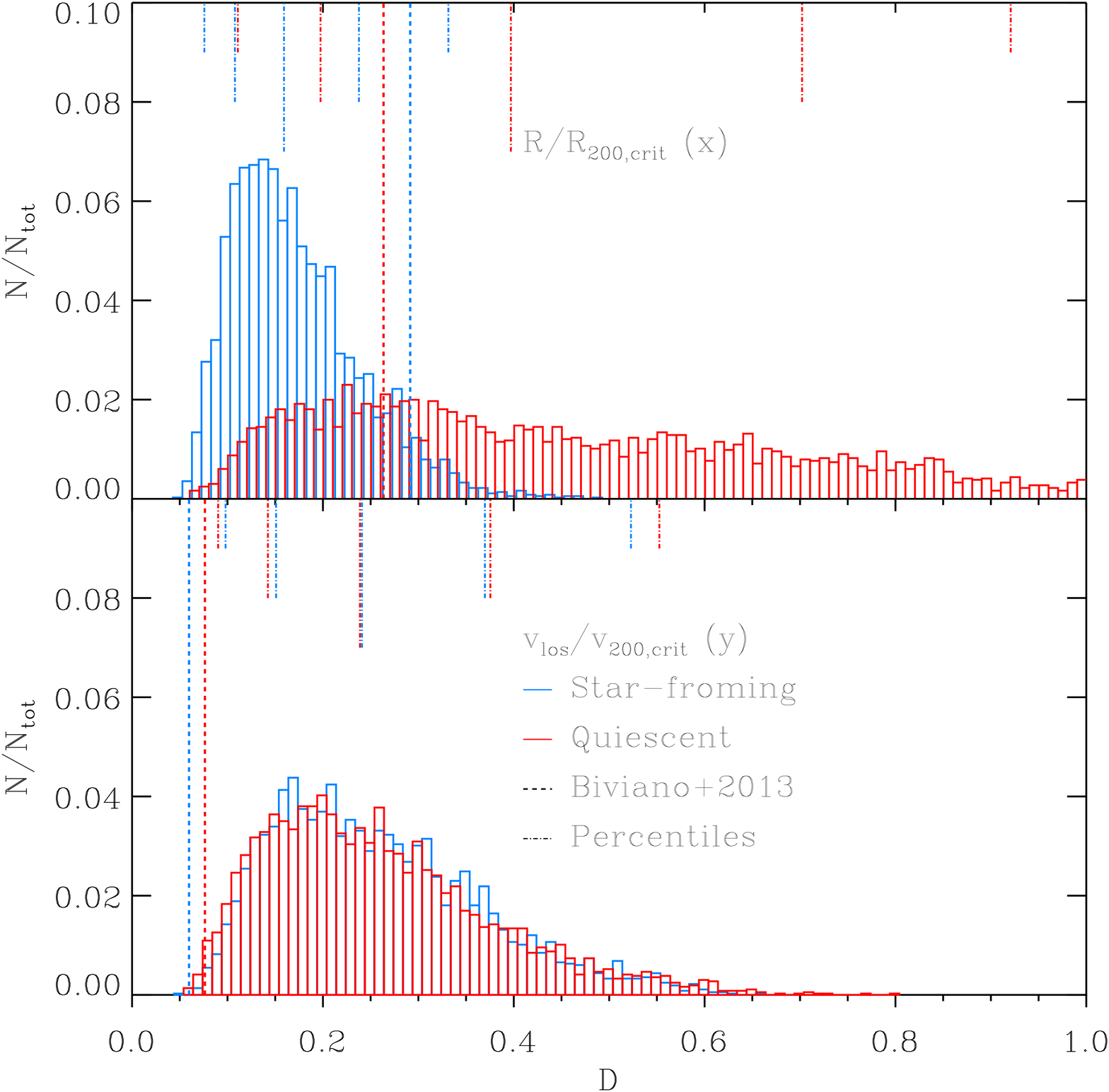}
    \caption{Histograms of the normalised frequency of the Kolmogorov-Smirnov (KS) statistic, quantifying the distance between cluster-cluster distribution functions in Magneticum. The top panel shows the Magneticum cluster-cluster histograms of the radial distribution of both star-forming (blue) and quiescent (red) cluster satellite galaxies as presented in Figure \ref{fig:ProjPS}. The bottom panel shows the same populations as the top panel, but along the line-of-sight velocity distribution. In each panel the dashed long vertical lines indicate the KS statistic from the comparison between \protect\cite{2013A&A...558A...1B} and the entire $86$ cluster Magneticum sample. The dash-dotted lines at the top of the panels show the percentiles of the Magneticum cluster-cluster histograms. Short lines correspond to the $2 \sigma$ region, medium lines correspond to the $1 \sigma$ region, and the longest line indicates the median of the distribution.}
    \label{fig:KShisto}
\end{figure}

\begin{figure}
    \centering
	\includegraphics[width=0.99\columnwidth]{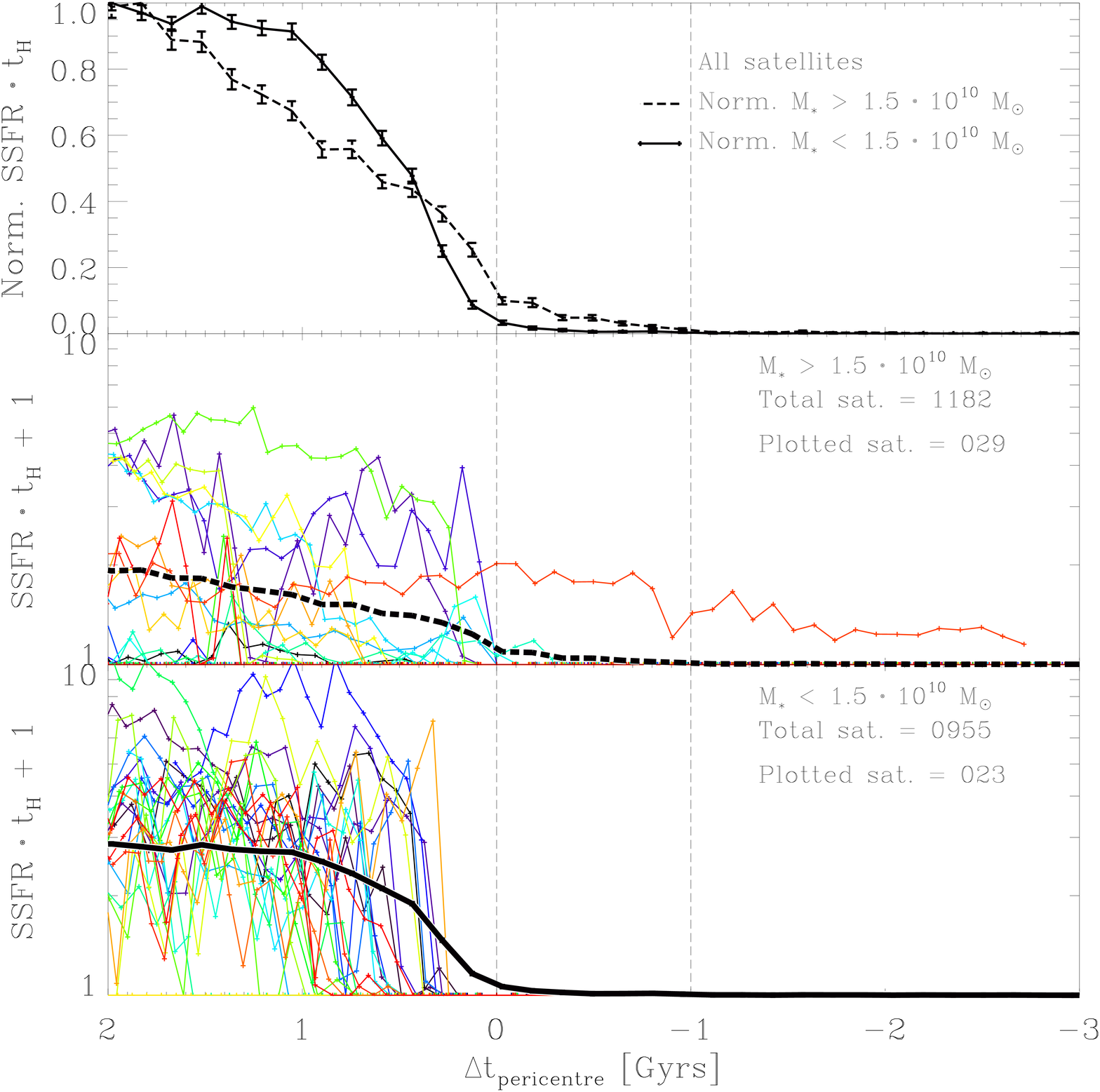}
    \caption{Same as \ref{fig:OrbitTrackBlue} but normalised to pericentre passage rather than cluster virial radius crossing, i.e. $\Delta t_{pericentre} = 0$ marks the pericentre passage of each satellite galaxy.}
    \label{fig:desc_peri}
\end{figure}

Further tests were conducted to distinguish whether: a.) simulated line-of-sight galaxy distributions are compatible with the observations (Figure \ref{fig:KShisto}) and b.) a correlation exists between pericentre passage and a star-burst (Figure \ref{fig:desc_peri}).

To compute the KS statistic shown in Figure \ref{fig:KShisto} each satellite galaxy distribution (both star-forming and quiescent and along each axis) of each cluster was compared with every other Magneticum cluster. The compared satellite galaxy sample is based on the same $86$ clusters as shown in Figure \ref{fig:ProjPS} and provides a statistical metric to access the similarity between individual Magneticum clusters.
As described in Section \ref{sub:LOS_PS}, we find that the differences between the single cluster observation by \cite{2013A&A...558A...1B} and the stacked Magneticum clusters is typically much less than the expected cluster to cluster variation within the simulation. The only exception to this is the radial distribution of star-forming galaxies, which, however, still lie well within the 2 $\sigma$ region of the cluster by cluster variation.

In Figure \ref{fig:desc_peri}, we investigated whether a correlation between pericentre passage of cluster satellite galaxies and a star-burst exists. We found no evidence for a star-burst. As we find a star-burst when normalising to cluster in-fall rather than pericentre passage (see Figure \ref{fig:OrbitTrackBlue}), this suggests that the star-burst is linked to in-fall rather than the pericentre passage, likely indicating the onset of ram-pressure stripping.

\bibliographystyle{mnras}


\bibliography{research}

\bsp	
\label{lastpage}
\end{document}